\documentclass{aa}
\usepackage{subcaption}
\usepackage[varg]{txfonts}
\usepackage{graphicx}
\usepackage{verbatim}
\usepackage{gensymb}
\usepackage{epstopdf}
\graphicspath{{./Figures/}}
\makeatletter
\newcommand{\RNum}[1]{\uppercase\expandafter{\romannumeral #1\relax}}
\begin{document}

\title{Sunspot rotation. \RNum{1}. A consequence of flux emergence}

\author{Z. Sturrock
  \and A. W. Hood
  \and V. Archontis
  \and C. M. McNeill}

\institute{School of Mathematics and Statistics, University of St Andrews, St Andrews, Fife, KY16 9SS, UK\\
email: \texttt{zoe.sturrock@st-andrews.ac.uk} } 

\date{Received 12 May 2015 / Accepted 2 August 2015}

\abstract {Solar eruptions and high flare activity often accompany the rapid rotation of sunspots. The study of sunspot rotation and the mechanisms driving this motion are therefore key to our understanding of how the solar atmosphere attains the conditions necessary for large energy release.} {We aim to demonstrate and investigate the rotation of sunspots in a 3D numerical experiment of the emergence of a magnetic flux tube as it rises through the solar interior and emerges into the atmosphere. Furthermore, we seek to show that the sub-photospheric twist stored in the interior is injected into the solar atmosphere by means of a definitive rotation of the sunspots.}{A numerical experiment is performed to solve the 3D resistive magnetohydrodynamic (MHD) equations using a Lagrangian-Remap code. We track the emergence of a toroidal flux tube as it rises through the solar interior and emerges into the atmosphere investigating various quantities related to both the magnetic field and plasma. }{Through detailed analysis of the numerical experiment, we find clear evidence that the photospheric footprints or sunspots of the flux tube undergo a rotation. Significant vertical vortical motions are found to develop within the two polarity sources after the field emerges. These rotational motions are found to leave the interior portion of the field untwisted and twist up the atmospheric portion of the field. This is shown by our analysis of the relative magnetic helicity as a significant portion of the interior helicity is transported to the atmosphere. In addition, there is a substantial transport of magnetic energy to the atmosphere. Rotation angles are also calculated by tracing selected fieldlines; the fieldlines threading through the sunspot are found to rotate through angles of up to $353\degree$ over the course of the experiment. We explain the rotation by an unbalanced torque produced by the magnetic tension force, rather than an apparent effect.}{}
\keywords{Sun: magnetic fields -- magnetohydrodynamics (MHD) -- methods: numerical}

\maketitle

\section{Introduction}

 Sunspot rotation has attracted the interest of many researchers over the years from both an observational and modelling perspective. The wide scope in observations of sunspot rotation merits a study of the mechanisms driving this motion. Coronal mass ejections (CMEs) and solar flares are often related to the rapid rotation of sunspots. Hence, the study of sunspot rotation is crucial to our understanding of such explosive events on the Sun as a possible mechanism for allowing the corona to achieve the necessary conditions for high energy release.

Just over a century ago, whilst working at the Kodaikanal Solar Observatory,~\cite{Evershed1909} first discovered evidence of the rotation of sunspots based on spectral observations. Since this initial evidence, sunspot rotations have been analysed in numerous observational studies. In recent years, detailed case studies such as~\cite{Yan2007},~\cite{Yan2009}, and~\cite{Min2009} and statistical analyses from~\cite{Brown2003} and \cite{Yan2008} have investigated this phenomena. These studies, along with several others, have shown that sunspots can exhibit significant rotation of the order of several hundreds of degrees over a few days.~\cite{Yan2008} conducted a statistical study of rotating sunspots using TRACE (Transition Region And Coronal Explorer), Hinode, and MDI (Michelson Doppler Imager) magnetograms from 1996 to 2007, in which they individually analysed 2959 active regions. They found 182 significantly rotating sunspots within 153 active regions. This is equivalent to approximately 5\% of active regions harbouring rotating sunspots. On the other hand,~\cite{Brown2003} conducted a more detailed study of seven sunspots using white light TRACE data to find sunspot centres and track notable features over time to calculate rotation rates. Rotation angles of between $40\degree$ and $200\degree$ were observed over periods of three to five days, resulting in an average rotation rate of a few degrees per hour. Six of these rotating spots resulted in subsequent flaring activity and the energisation of the corona. 

Interestingly, measurements of sunspot rotation have given variable results depending on the methodology employed. For example,~\cite{Min2009} and~\cite{Yan2009} analysed the same active region, NOAA $10930$, and found notably different results.~\cite{Min2009} noted a counter-clockwise rotation of $540$ degrees over five days, whereas~\cite{Yan2009} focused on the X$3.2$ flare that followed the rapid rotation of $259$ degrees over four days. Furthermore, although~\cite{Brown2003} and~\cite{Yan2007} both concluded that different parts of sunspots often rotate at differing speeds,~\cite{Brown2003} noted that the highest rotation rate was found in the penumbra while~\cite{Yan2007} concluded that the highest rotation rate was found in the umbra. This suggests that sunspots do not necessarily rotate as a rigid body.~\cite{Yan2007} concluded that twist can be created by a variation in rotation rate with distance from the centre of a sunspot. This twist can then be injected into the corona kinking the magnetic loops and driving flare activity.

Eruptions and flares are, in fact, often correlated with rotational motions of sunspots. Observations have shown direct evidence of the energisation of the corona by these rotations. The initiation of CMEs by sunspot rotation was studied in detail by~\cite{Torok2014} from both an observational and modelling viewpoint. They found that the rotation of sunspots can significantly weaken the magnetic tension of the overlying field in active regions and can then trigger an eruption in this region. This is an alternative explanation for eruptions caused by rotation of sunspots to the common theory that eruptions are triggered by a direct injection of twist~\citep{Yan2007}.

This paper aims to discuss and simulate a possible mechanism for the rotation of sunspots. Two mechanisms were discussed in~\cite{Brown2003}, namely, photospheric flows and magnetic flux emergence. Photospheric flows are primarily due to the large scale effect of differential rotation and localised motions resulting from magneto-convective dynamics. The effects of differential rotation are kept to a minimum in~\cite{Brown2003} as the images were derotated prior to measuring the velocities. The second mechanism discussed was the emergence of magnetic flux.~\cite{Brown2003} suggested that the photospheric footpoints of a flux tube are observed to rotate as the tube emerges. This is the proposed mechanism for the rotation of sunspots which we will investigate. This mechanism seems the most appropriate according to the findings of~\cite{Brown2003} as well as a case study from~\cite{Min2009}. The latter study supports flux emergence as a viable mechanism as they discovered that the rotation speed increases in close relation to the growth of the sunspot of interest, which we attribute to the emergence of flux. The next logical question is, how can flux emergence drive sunspot rotation? Two possible explanations were suggested by~\cite{Min2009}. They conjectured that the rotation may be an apparent motion caused by a twisted flux tube rising vertically and the fieldlines successively crossing the photospheric boundary at different locations as a result of the the twisted structure of the field. In this case there is no real horizontal motion of the plasma; instead this rotation is a virtual effect caused by continued displacement of the footpoints. Alternatively,~\cite{Min2009} proposed that the observed rotation may represent the real horizontal motion of the plasma due to a net torque originating in the interior driving the plasma to rotate on the photospheric boundary. The torque will be examined later in an effort to determine its origin.

The process of magnetic flux emergence has been considered numerically in various experiments in recent years (for instance~\cite{Fan2001},~\cite{Archontis2004},~\cite{Hood2009} amongst many others). The widely accepted picture of sunspot formation is that an $\Omega$-shaped flux tube rises from the base of the convection zone until its apex intersects the photosphere to form a pair of sunspots. If a magnetic flux tube is in pressure balance and thermal equilibrium with its surroundings, the tube will be less dense than its surroundings, and will therefore be buoyant. This mechanism allows a flux tube to rise to the stably stratified photosphere where it remains until the tube is able to enter the atmosphere by initiation of a second instability, namely the magnetic buoyancy instability. A pair of concentrations of opposite polarities, known as bipolar sunspots, mark the intersection of the field at the surface. In these experiments, the computational domain typically models a region from the top of the solar interior to the lower corona. A magnetic flux tube is then placed in the solar interior and is made buoyant by either introducing a density deficit or imposing an initial upward velocity. A recent review of simulations relating to this emergence process was undertaken by~\cite{Hood2012}.

Investigations of magnetic flux emergence as a possible mechanism for sunspot rotation have been conducted in the past.~\cite{Longcope2000} suggested that the rotation of sunspots is a consequence of the transport of helicity from the convection zone to the corona as a twisted flux tube emerges. They also demonstrated that a torsional Alfv\'{e}n wave will propagate downwards at the instance of emergence due to a mismatch in current between the highly twisted interior and stretched coronal portion of the field. In addition,~\cite{Gibson2004} explained the rotation as an observational consequence of the emergence of a flux tube through the photosphere. An investigation of the transport of magnetic energy and helicity in an emerging flux model was conducted by~\cite{Magara2003}. They found that emergence generates horizontal flows as well as vertical flows, both of which contribute to the injection of energy and helicity to the atmosphere. The contributions by vertical flows are dominant initially but horizontal flows are the primary source at a later stage. A more comprehensive study of the horizontal flows at the photosphere during emergence was later performed by~\cite{Magara2006} in which they found that rotational flows formed in each of the polarity concentrations soon after the intersecting flux tubes became vertical. Furthermore, the rotational movement of sunspots in a $3$D MHD simulation has been examined by~\cite{Fan2009} where significant vortical motions developed as a torsional Alfv\'en wave propagated along the tube.~\cite{Fan2009} noted that the rotation of the two polarities twisted up the inner fieldlines of the emerged field, thereby transporting twist from the interior portion of the flux tube to the stretched coronal loop. In this simulation, a cylindrical flux tube is inserted into the solar interior using the cylindrical model developed by~\cite{Fan2001}.

~\cite{Hood2009} perform a complementary simulation to that of~\cite{Fan2001}, replacing the initial cylindrical flux tube with a toroidal tube. A common shortcoming of the cylindrical model in simulations without convective flows is that the axis of the tube never fully emerges. Altering the geometry of the flux tube to a curved shape allows for the axis of the tube to rise into the corona. Current theories suggest that the Sun's magnetic field is created in the solar interior by dynamo action. Hence, a half-torus shaped flux tube imitates a field anchored deeper within the solar interior. The rotation of sunspots has not yet been investigated using this toroidal model; this is what we aim to study. 

In the present paper, we perform a resistive $3$D MHD simulation of a toroidal flux tube placed in the solar interior and track its emergence through the photosphere and lower atmosphere. We study the role of rotational flows at the photosphere while also investigating the transport of magnetic helicity and energy. In addition, we explicitly calculate the angles of rotation in our experiment to directly compare with observations. Our main aim is to demonstrate that the interior magnetic field is untwisting as the tube emerges causing an injection of twist into the atmosphere as well as a rotation of the sunspots. This will be accomplished by performing a detailed study investigating quantities relating to both the magnetic field and plasma.

The remainder of the paper is structured as follows. In Section~\ref{sec:modelsetup}, we describe the MHD equations used in our model as well as outlining the numerical approach employed. We also describe the initial setup of our experiment, detailing both the initial atmosphere and the model of the sub-photospheric magnetic field inserted in the solar interior. In Section~\ref{sec:results}, the simulation results are presented, with emphasis on the rotational motions that develop within the two polarities on the photospheric plane. Our analysis also includes an investigation of the sunspot rotation angle, the flow vorticity at the photosphere, the current, relative magnetic helicity, magnetic energy and twist. Finally, in Section~\ref{sec:conclude} we summarise the conclusions and discuss the implications of this simulation and future projects that stem from this work.

\section{Model setup}
\label{sec:modelsetup}

In this section, we outline the numerical setup of our experiment, detailing the initial atmospheric and magnetic field configuration.

\subsection{Model equations and numerical approach}
For the experiment performed in this paper, we numerically solve the $3$D time-dependent resistive MHD equations, as described below in non-dimensionalised form:
\begin{eqnarray}
\frac{\mathrm{D} \rho}{\mathrm{D} t} & = & - \rho \nabla \cdot \mathbf{v},\\
\frac{\mathrm{D}\mathbf{v}}{\mathrm{D}t} & = & \frac{1}{\rho}(\mathbf{\nabla} \times \mathbf{B}) \times 
\mathbf{B} - \frac{1}{\rho}\mathbf{\nabla} p - g\mathbf{\hat{z}}+\frac{\nabla \cdot \mathbf{T}}{\rho},\\
\frac{\mathrm{D} \mathbf{B}}{\mathrm{D} t} & = & (\mathbf{B} \cdot \nabla)\mathbf{v} - \mathbf{B}(\nabla \cdot \mathbf{v}) - \nabla \times (\eta \mathbf{\nabla} \times \mathbf{B}),\\
\frac{\mathrm{D} \epsilon}{\mathrm{D}t} & = & - \frac{p}{\rho}\mathbf{\nabla} \cdot \mathbf{v} + \frac{\eta}{\rho}j^2 + \frac{Q_{\text{visc}}}{\rho},
\end{eqnarray}
with specific energy density,
\begin{equation}
\epsilon = \frac{p}{(\gamma-1)\rho},
\end{equation}
and the electric current density,
\begin{equation}
\mathbf{j} = \bf \nabla \times \mathbf{B}.
\end{equation}
The basic dimensionless quantities used in these equations are the density $\rho$, time $t$, velocity $\mathbf{v}$, magnetic field $\mathbf{B}$, pressure $p$, gravity $g$, electric field $\mathbf{E}$, specific energy density $\epsilon$, resistivity $\eta$ and temperature $T$. The ratio of specific heats, $\gamma$, is taken to be $5/3$. The viscosity tensor is denoted by $\mathbf{T}$ and the contribution of viscous heating to the energy equation is represented by $Q_{\textbf{{visc}}}$. The difference between simulations with and without a small viscosity are found to be negligible. The variables are made dimensionless against photospheric values. These values are: pressure, $p_{\text{ph}} = 1.4 \times 10^4$ Pa; density, $\rho_{\text{ph}} = 3 \times 10^{-4}$ kg m$^{-3}$; temperature $T_{\text{ph}} = 5.6 \times 10^3$ K; pressure scale height, $H_{\text{ph}} = 170$ km; velocity, $v_{\text{ph}} = 6.8$ km s$^{-1}$; time, $t_{\text{ph}} = 25$ s and magnetic field, $B_{\text{ph}} = 1300$ Gauss. The dimensionless resistivity $\eta$ has been taken as uniform set at a value of $0.005$ in our experiment. All quantities from now on will be dimensionless, unless units are provided. In order to reach the true quantities with physical units, the dimensionless quantities should be multiplied by their photospheric values as given above.

The code used to simulate the emergence process is a 3D Lagrangian remap code, \emph{Lare3d}~\citep{Arber2001}. The code uses a staggered grid and is second order accurate in both space and time. The LARE code can be divided into two main steps; the Lagrangian step and the remap step. In the Lagrangian step the equations are solved in a frame that moves with the fluid. This causes the grid to be distorted, so in order to put the variables back on to the original (Eulerian) grid, a remap step is used. The code accurately resolves shocks by using a combination of artificial viscosity and Van Leer flux limiters. This code also includes a small shock viscosity to resolve shocks and the associated shock heating term in the energy equation. 

The equations are solved on a uniform Cartesian grid ($x$, $y$, $z$) comprised of $512^3$ grid points. The box spans from $-50$ to $50$ in the $x$ and $y$ directions and from $-25$ to $75$ in the $z$ direction in dimensionless units. This corresponds to a physical size of $17$ Mm$^3$. The boundary conditions are periodic on the side walls of the computational domain and the top and bottom boundaries are closed with $\mathbf{v}=\mathbf{0}$. For all other variables we set the normal derivatives to zero. As a consequence, the magnetic field is fixed on the bottom boundary.

\subsection{Initial atmosphere}
\label{subsec:atmos}
\begin{figure}[ht]
\centering
\includegraphics[width=8cm]{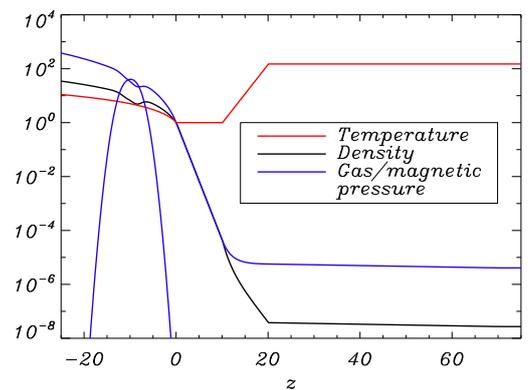}
\caption{Initial stratification of model atmosphere. The initial profiles are plotted on a log scale against height where red denotes the temperature distribution, black denotes the density and blue represents the magnetic pressure in the solar interior and the gas pressure throughout the domain.}
\label{fig:rotinitialatmos}
\end{figure}

The initial stratification of the atmosphere is the same as many previous flux emergence studies, for example~\cite{Hood2009}. The computational domain is split into four regions: the solar interior ($z<0$); the photosphere/chromosphere $0 \le z < 10$;  the transition region $10 \le z < 20$ and the lower corona $z \ge 20$. The solar surface is taken to be the plane at $z=0$. The stratification is uniform across the horizontal plane and varies only with height $z$. The solar interior is taken to be marginally stable to convection by assuming constant entropy in this region. This assumption seems appropriate as the focus of this experiment is the evolution of the magnetic field. The photosphere/chromosphere is taken as an isothermal region with a dimensionless temperature of unity by design. The temperature distribution in the transition region is a power-law profile to model the steep temperature gradient observed here. Lastly, the isothermal corona is set at a temperature $150$ times that of the photosphere ($T_{\text{cor}} = 150$). To summarise, the non-dimensionalised temperature profile is specified as
\[T(z) = \left\{
  \begin{array}{cr}
    1 - \frac{(\gamma-1)}{\gamma}z &  z < 0,\\
    1 & 0 \le z < 10 ,\\
    150^{\left(\frac{z-10}{10}\right)} & 10 \le z < 20,\\
    150 & z \ge 20.
  \end{array}
\right.
\]
The pressure and density are then calculated by numerically solving the dimensionless hydrostatic balance equation $\frac{\mathrm{d}p}{\mathrm{d}z} = - \rho g$ and the dimensionless ideal gas law. The resulting logarithms of the initial temperature, density and pressure of the stratified background are shown in Figure~\ref{fig:rotinitialatmos}.

\subsection{Initial magnetic field}
We choose to leave the atmosphere unmagnetised by neglecting an ambient field and concentrating on the sub-photospheric field. We insert a magnetic flux tube into the hydrostatic solar interior. As we want the setup to remain in equilibrium, and the tube to be in force balance, we require
\begin{equation}
-\mathbf{\nabla} p + \mathbf{j} \times \mathbf{B}+ \rho \mathbf{g} = \mathbf{0}.\nonumber
\end{equation}
Given that the background environment is in hydrostatic pressure balance, this reduces to
\begin{equation}
\mathbf{\nabla} p_{\text{exc}} = \mathbf{j} \times \mathbf{B},\label{eqrot:hydro}
\end{equation}
where $p_{\text{exc}}$ is the pressure excess such that the gas pressure in the tube, $p_t$, is defined to be $p_b+p_{\text{exc}}$ where $p_b$ is the background gas pressure. Explicitly, we have split the gas pressure into a background pressure component that balances gravity and a second component that balances the Lorentz force. The next obstacle is to choose the form of magnetic field to prescribe in the solar interior. The interior magnetic field cannot be observed therefore we choose simple models for the sub-photospheric field to initiate emergence. We choose to place a toroidal flux tube with twisted fieldlines in the solar interior. The toroidal model we utilise was derived fully in~\cite{Hood2009}. Compared with the standard choice of a cylindrical tube, this models the emergence of the top part of an $\Omega$-shaped loop that is rooted much deeper in the solar interior. Here, we simply outline the main steps of the derivation.

First, we express our original cartesian coordinates ($x, y, z$) in terms of cylindrical coordinates $(R, \phi, x$), given that $y$ is the coordinate describing the direction of the axis of the tube, and $z$ denotes the height from the solar surface as previously. We note that
\begin{equation}
R^2=y^2+(z-z_{\text{base}})^2,\quad\text{with }y=R\cos{\phi}\text{ and }z-z_{\text{base}}=R\sin{\phi},\nonumber
\end{equation}
where $z_{\text{base}}$ is the value of $z$ at the base of the computational domain. The magnetic field is then expressed, in terms of the flux function $A=A(R,x)$, as
\begin{eqnarray}
\mathbf{B} & = & \mathbf{\nabla} A \times \mathbf{\nabla} \phi + B_{\phi}\mathbf{e}_{\phi}\nonumber\\
 & = & -\frac{1}{R}\frac{\partial A}{\partial x}\mathbf{e}_R + B_{\phi}\mathbf{e}_{\phi} + \frac{1}{R}\frac{\partial A}{\partial R}\mathbf{e}_{x},\nonumber
\end{eqnarray}
where $A$ is constant along magnetic fieldlines. In this derivation, we assume that the magnetic field is rotationally invariant, i.e. independent of $\phi$. We note that this form of magnetic field automatically satisfies the solenoidal constraint ($\nabla \cdot \mathbf{B} = 0$). 
Inserting the magnetic field, $\mathbf{B}$, in terms of the flux function, $A$, into equation~\eqref{eqrot:hydro}, noting $p_{\text{exc}} = F(A(R,x))$ and $RB_{\phi}=G(A(R,x))$ along with some algebraic manipulation yields,
\begin{equation*}
-\frac{1}{R}\left\{\frac{1}{R}\frac{\partial^2A}{\partial x^2} + \frac{\partial}{\partial R}\left(\frac{1}{R}\frac{\partial A}{\partial R}\right)\right\}\mathbf{\nabla} A - \frac{1}{R^2}b_{\phi}\frac{db_{\phi}}{dA}\mathbf{\nabla} A = \frac{dp_{\text{exc}}}{dA}\mathbf{\nabla} A,
\label{eq:rotgradshafranov}
\end{equation*}
where $b_{\phi}=RB_{\phi}$. We note that all the terms are in the direction of $\nabla A$, therefore the \emph{Grad-Shafranov} equation is of the form
\begin{equation}
R\frac{\partial}{\partial R}\left(\frac{1}{R}\frac{\partial A}{\partial R}\right)+\frac{\partial^2A}{\partial x^2} + b_{\phi}\frac{db_{\phi}}{dA}+R^2\frac{dp_{\text{exc}}}{dA} = 0.\label{eq:rotgradshaf2}
\end{equation}

Following the strategy used by~\cite{Hood2009}, we now convert our system to a local toroidal coordinate system ($r,\theta,\phi$), such that
\begin{eqnarray}
r^2=x^2+(R-R_0^2),\quad\text{with }R-R_0=r\cos{\theta}\text{ and }x=-r\sin{\theta},\nonumber
\end{eqnarray}
where $R_0$ is the major axis of the toroidal loop and $a$ is the minor radius of the toroidal loop. Equation~\eqref{eq:rotgradshaf2} can be re-written in these local coordinates by assuming $a<<R_0$ and in turn $r<<R_0$, i.e. assuming that the minor radius of the torus is much smaller than the major radius. We can expand the solution in powers of $a/R_0$ such that to leading order, equation~\eqref{eq:rotgradshaf2} becomes

\begin{equation}
\frac{B_{\theta}}{r}\frac{d}{dr}(rB_{\theta}) + \frac{1}{2}\frac{d(R^2B_{\phi}^2/R_0^2)}{dr}+\frac{dp_{\text{exc}}}{dr} = 0.\nonumber
\end{equation}
This has exactly the same form as the standard cylindrical equation found in~\cite{Archontis2004}. We can therefore choose the solutions to be the same as the cylindrical flux tube. Specifically,
\begin{eqnarray}
B_{\phi}=\frac{R_0}{R}B_0e^{-r^2/a^2}\quad\text{and}\quad B_{\theta}=\alpha rB_{\phi} = \alpha\frac{R_0}{R} B_0re^{-r^2/a^2},\nonumber
\end{eqnarray}
where $B_0$ is the axial field strength and $\alpha$ is the twist. The local toroidal field can also be approximated to $O (a/R_0)$. Again using $R=R_0+r\cos{\theta} \sim R_0 + O(a/R_0)$
this approximates the local toroidal field to
\begin{equation}
B_{\phi} \sim B_0e^{-r^2/a^2}\quad\text{and}\quad B_{\theta} \sim \alpha B_0re^{-r^2/a^2}.\nonumber
\end{equation}
With these approximations to the local toroidal field, the pressure can be calculated by exact comparison with the cylindrical model,
\begin{equation}
p_{\text{exc}}(r) = \frac{B_0^2}{4}e^{-2r^2/a^2}(\alpha^2a^2 - 2\alpha^2r^2 -2).\nonumber
\end{equation}
This balances equation~\eqref{eqrot:hydro} and forces the flux tube to sit in equilibrium. However, to initiate the emergence we introduce a density deficit as given by
\begin{equation}
\rho_{\text{def}}(r) = \frac{B_0^2}{4T(z)}e^{-2r^2/a^2}(\alpha^2a^2 - 2\alpha^2r^2 -2).\nonumber
\end{equation}
using the temperature profile specified in Section~\ref{subsec:atmos}. This makes the flux tube lighter than its surroundings and allows it to begin rising.

Lastly, we need to re-express the magnetic field in terms of cartesian coordinates as this is the form that we require for the input of our MHD simulations. Before we can do this directly, we must first express the field in cylindrical coordinates ($R,\phi,x)$ as 
\begin{eqnarray}
B_R &=& - B_{\theta}(r)\sin{\theta} =  B_{\theta}(r)\frac{x}{r},\nonumber\\
B_x & = & - B_{\theta}(r)\cos{\theta} = - B_{\theta}(r)\frac{R-R_0}{r}.\nonumber
\end{eqnarray}
Similarly, this cylindrical field can then be converted into cartesian coordinates to set the magnetic field as
\begin{eqnarray}
B_x  & = & - B_{\theta}(r)\frac{R-R_0}{r},\nonumber\\
B_y & = & - B_{\phi}(r)\frac{z-z_{\text{base}}}{R} + B_{\theta}(r)\frac{x}{r}\frac{y}{R},\nonumber\\
B_z & = & B_{\phi}(r) \frac{y}{R} + B_{\theta}(r)\frac{x}{r}\frac{z-z_{\text{base}}}{R},\nonumber
\end{eqnarray}
where 
\begin{eqnarray}
B_{\phi} = B_0e^{-r^2/a^2}\quad\quad\text{and}\quad B_{\theta} = \alpha r B_{\phi} = \alpha B_0re^{-r^2/a^2}.\nonumber
\end{eqnarray} 
We note that this corrects an error in~\cite{Hood2009} where the sign of $B_R$ and $B_x$ were interchanged. Fortunately, this error was not significant as it is equivalent to using an initial twist, $\alpha$, of the opposite sign.

\subsection{Parameter choice}

\begin{figure}[ht]
\centering
\includegraphics[scale=0.27]{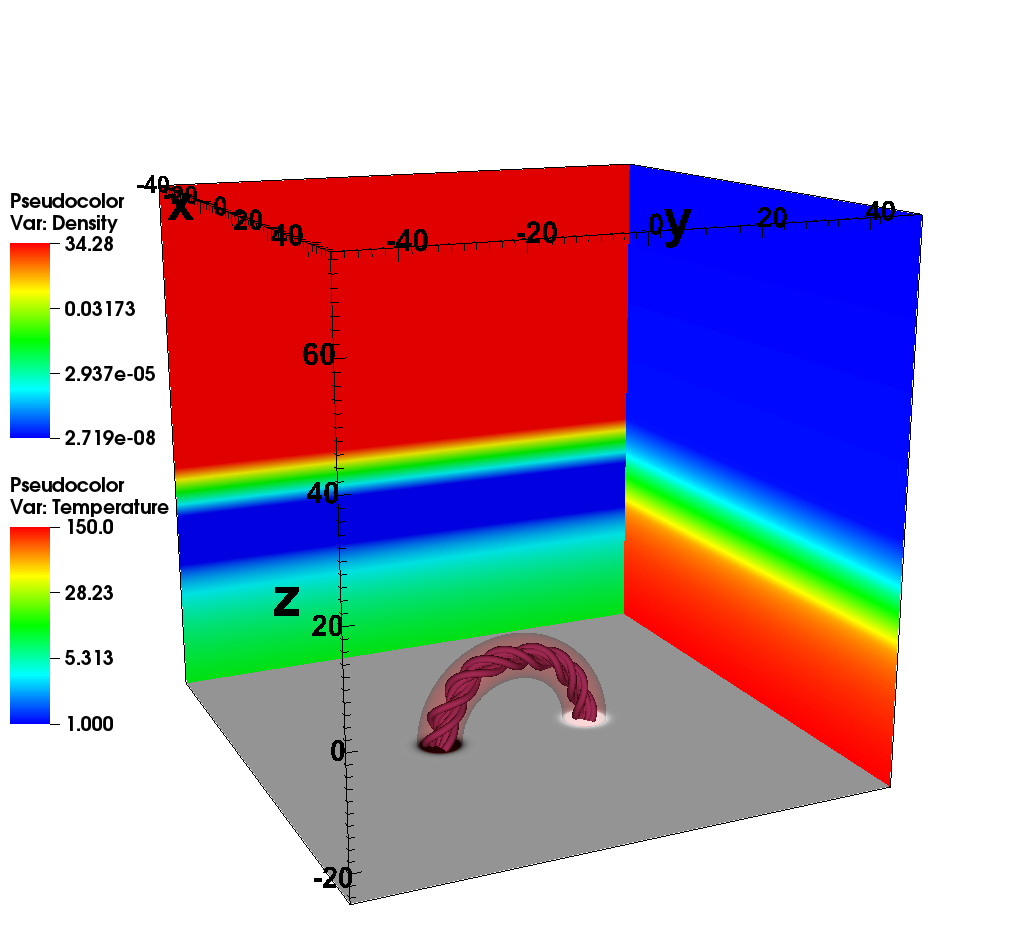}
\caption{Summary of initial setup. The background density distribution is shown on the right wall, the temperature distribution on the back wall, a selection of fieldlines are shown in purple and an isosurface of the magnetic field ($|\mathbf{B}|=1$) is overplotted.}
\label{fig:rotinitialsetup}
\end{figure}
In this experiment, we set the magnetic field strength at the apex of the tube as $B_0 = 9$ ($11700$~G) and the twist as $\alpha = 0.4$ (right-hand twisted). This means that each fieldline undergoes an angle of $0.4~$rad per unit distance, and is equivalent to approximately three full turns of twist in the initial sub-photospheric field. The base of the computational domain is set at $z=-25$. The major radius of the torus is $R_0=15$ and the minor radius is $a=2.5$. The total flux threading a cross section of the tube is $6.6\times10^{19}$~Mx, typical of a small active region or a large ephemeral region. The initial set-up of the experiment is summarised in Figure~\ref{fig:rotinitialsetup}.

\section{Analysis}
\label{sec:results}
In order to examine and quantify the rotation of the sunspots in our simulation, we undertake a detailed investigation of a number of quantities relating to both the plasma and magnetic field in an attempt to demonstrate the untwisting of the interior field and rotational flows at the photosphere. First of all, we briefly describe the overall evolution of the field detailing the main phases of emergence.

\subsection{Evolution of magnetic field}
The flux tube begins to rise buoyantly to the photosphere due to the initial density deficit introduced and continues to rise because of the decreasing temperature stratification in the interior. However, at the photosphere, the plasma is stably stratified with a constant temperature and the flux tube is no longer buoyant. Therefore, the magnetic field finds another way to rise and expand into the corona, namely the magnetic buoyancy instability. In order to initiate this instability, a criterion must be satisfied as derived by~\cite{Acheson1979}. Typically, the onset of this instability occurs when the plasma $\beta$ drops to one. At this stage, the magnetic pressure exceeds the gas pressure and the field expands into the atmosphere. See~\cite{Archontis2004} and~\cite{Hood2012} for a full description of this instability.

\begin{figure}[ht]
\centering
\begin{subfigure}{0.48\textwidth}
\centering
\includegraphics[width=6cm]{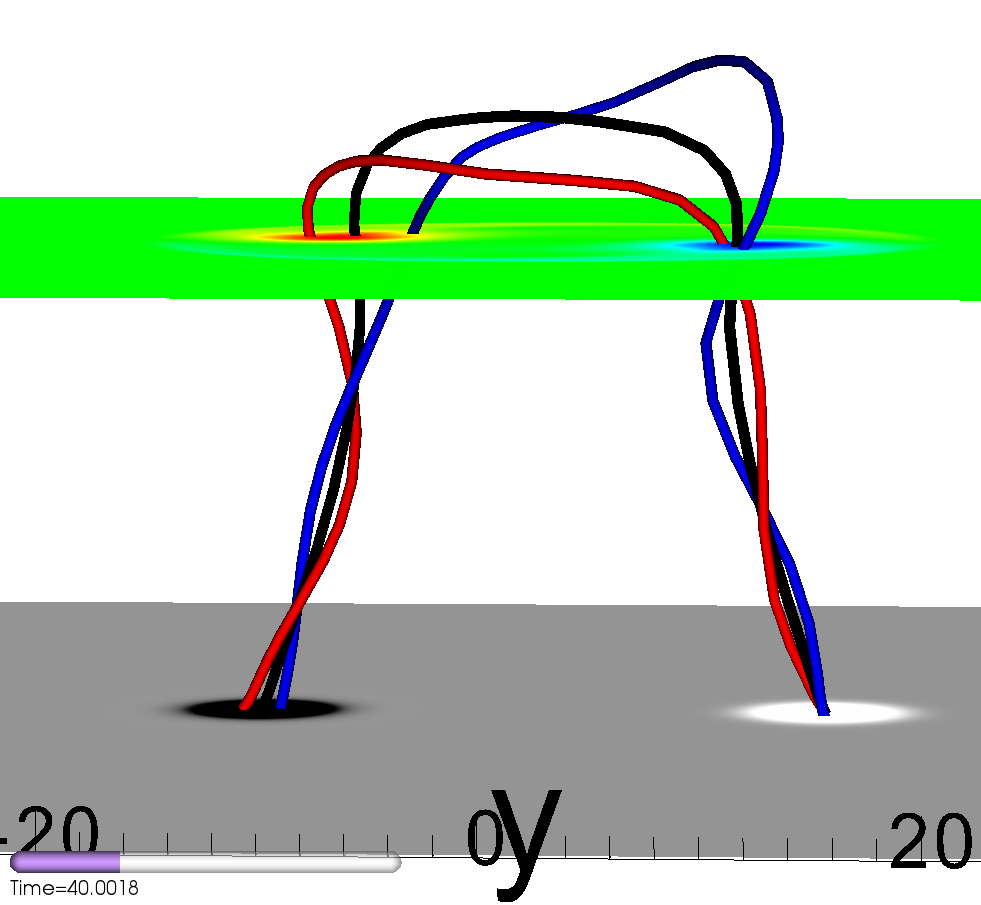}
\subcaption{$t=40$}
\label{fig:rotinteriorfieldt40}
\end{subfigure}\\
\begin{subfigure}{0.48\textwidth}
\centering
\includegraphics[width=6cm]{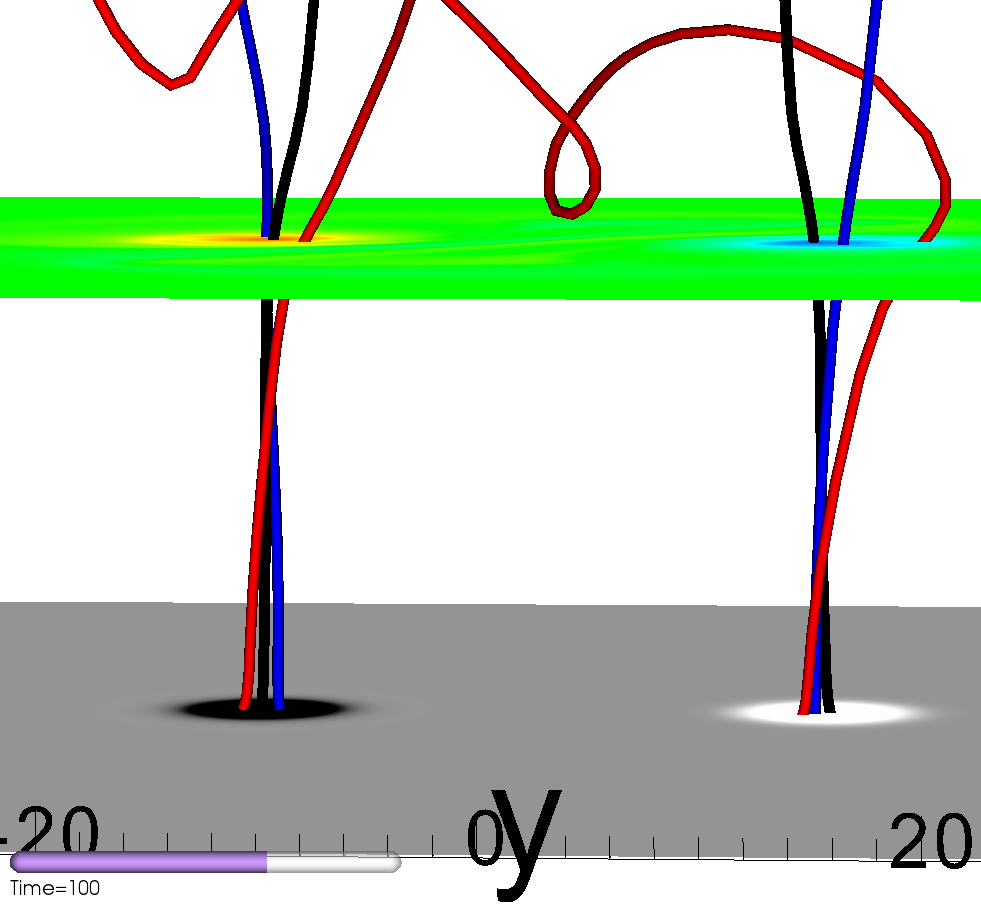}
\subcaption{$t=100$}
\label{fig:rotinteriorfieldt100}
\end{subfigure}
\caption{Visualisation of the field in the interior at times $t=40$ and $t=100$ respectively as traced from the lower negative footpoint (left). A movie of this figure is included in the electronic version.} 
\label{fig:rotinteriorfield}
\end{figure}

In Figure~\ref{fig:rotinteriorfield}, we have included two figures illustrating the evolution of the magnetic fieldlines as the flux tube emerges concentrating on the interior portion of the field. We have traced three fieldlines from $(0,-14,-25)$ (blue), $(0,-15,-25)$ (black) and $(0,-16,-25)$ (red) respectively. Initially the flux tube has $3$ full turns of twist in total. At $t=40$, the flux tube reaches the photosphere and the legs start to straighten. At this time, there is an emerged section of flux tube with about half a turn of twist that will subsequently expand into the corona. However, there is still a considerable amount of twist submerged, approximately a full twist in each leg. In the subsequent evolution there is a definitive unwinding of the submerged twist leading to a final state in which there is virtually no twist in the interior portion of the tube as evidenced by the visualisation of the field at $t=100$..
 We suggest that this may be governed by torsional Alfv\'en waves. The release of sub-photospheric twist as a mechanism for the propagation of a torsional Alfv\'en wave is investigated later.

\subsection{Driver of rotational motion}
\begin{figure}[ht]
\centering
\includegraphics[scale=0.4]{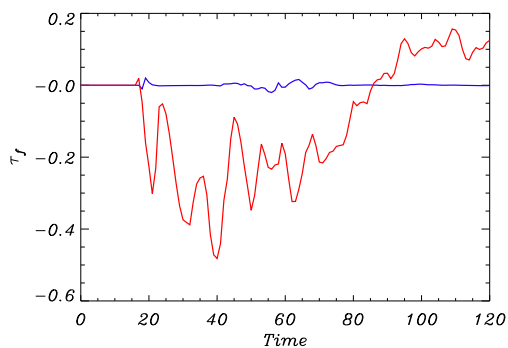}
\caption{Surface integral of torque due to the magnetic tension force (red) and the magnetic pressure force (blue) within a circular contour of radius $2.5$ around a point $P$ corresponding to the maximum of the vertical magnetic field.}
\label{fig:rottorque}
\end{figure}

Before we examine various quantities at the photosphere in an effort to demonstrate the untwisting of the interior field, we discuss the underlying reason for a rotational movement at the photosphere when a twisted magnetic structure emerges. The basic cause for this rotational motion is the behaviour of the Lorentz force. Following the explanation given by~\cite{Cheung2014}, we consider a circular closed curve lying on the photospheric plane enclosing some point $P$ denoting the location of the maximum of $B_z$. We note that the torque is the tendency of a force to rotate an object about an axis, and is given by $\displaystyle{\mathbf{T}=\mathbf{r} \times \mathbf{F}}$ where $\mathbf{r}$ is the displacement vector and $\mathbf{F}$ is the given force. If we consider the torque due to the various forces acting on the plasma and magnetic field through the surface confined by this closed contour, we find that the magnetic tension is the only force that provides a net torque. That is, the torque contributions from the magnetic pressure and gas pressure forces through this surface vanish and any non-zero torque is a result of the magnetic tension. Explicitly,
\begin{eqnarray}
\tau_p & = & \iint{\mathbf{r} \times \mathbf{\nabla} (-p_{\text{gas}}) \cdot \text{\bf dS}} = 0,\nonumber\\
\tau_{mp} & = & \iint{\mathbf{r} \times \mathbf{\nabla} \left(-\frac{B^2}{2\mu}\right) \cdot \mathrm{\bf dS}} = 0,\nonumber\\
\tau_{mt} & = & \iint{\mathbf{r} \times \left(\frac{1}{\mu}(\mathbf{B} \cdot \nabla) \mathbf{B}\right)   \cdot \mathrm{\bf dS}},\nonumber
\end{eqnarray}
where $\mathbf{r}$ is the displacement vector of a point on the curve from $P$. This is in fact a general result that can be proved for any force of the form $\mathbf{F} = \mathbf{\nabla} f$. To demonstrate this we have calculated the torque due to the magnetic tension and magnetic pressure within a circular contour of radius $a$ ($2.5$) surrounding the location of the maximum of $B_z$, as displayed in Figure~\ref{fig:rottorque}. In this case, it is clear that there is no contribution from the magnetic pressure force. Hence, we speculate that the driving motion of the rotation at the photosphere may be governed by the unbalanced torque produced by the magnetic tension force. This is characteristic of an Alfv\'{e}n wave.

\subsection{Rotation angle}

\begin{figure}[ht]
\centering
\begin{subfigure}{0.24\textwidth}
\includegraphics[width=0.9\textwidth]{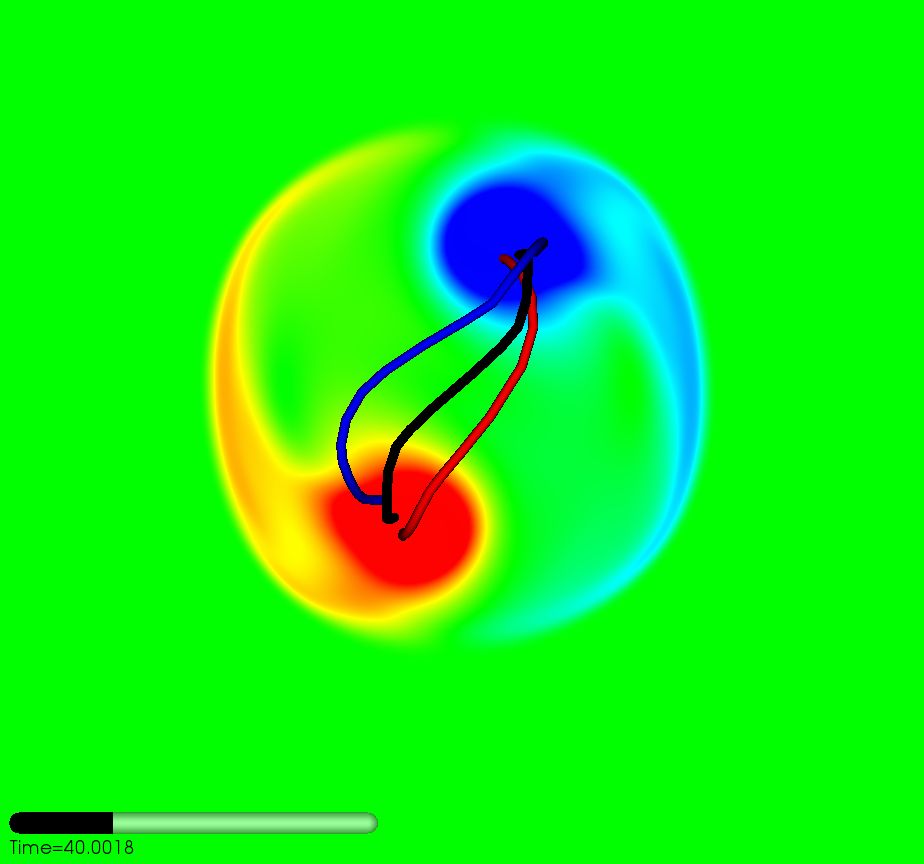}
\subcaption{$t=40$}
\label{fig:rotsunspotsfielda}
\end{subfigure}~
\begin{subfigure}{0.24\textwidth}
\includegraphics[width=0.9\textwidth]{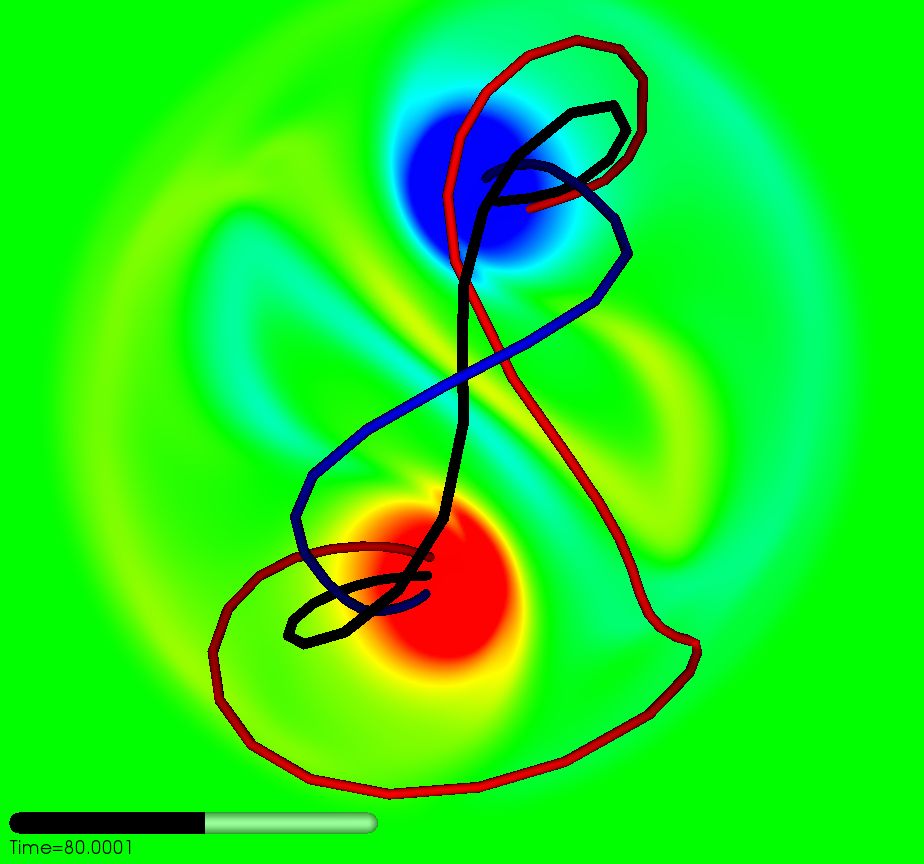}
\subcaption{$t=80$}
\label{fig:rotsunspotsfieldb}
\end{subfigure}
\caption{Visualisation of the axis of the flux tube (black fieldline) as well as two fieldlines (red and blue) spaced either side of the axis for comparison at selected times. A movie of this figure is included in the electronic version.}
\label{fig:rotsunspot}
\end{figure}

As our previous analyses suggest that the interior portion of the flux tube is untwisting and significant rotational motions develop within the sunspots, we have again traced three fieldlines from the base to the photosphere in an attempt to visualise this motion and quantify the amount by which the fieldlines have rotated. The axis of the flux tube has been traced from the lower negative footpoint as well as two fieldlines either side of the axis in the $y-$direction, i.e. we have traced fieldlines from $(0,-14.5,-25)$ (blue), $(0,-15,-25)$ (black) and $(0,-15.5,-25)$ (red).  A schematic of the traced fieldlines is shown in Figure~\ref{fig:rotsunspotsfielda} and Figure~\ref{fig:rotsunspotsfieldb} for times $t=40$ and times $t=80$ respectively. 
\begin{figure}\hspace{-0.5cm}
\centering
\begin{subfigure}{0.26\textwidth}
\includegraphics[width=\textwidth]{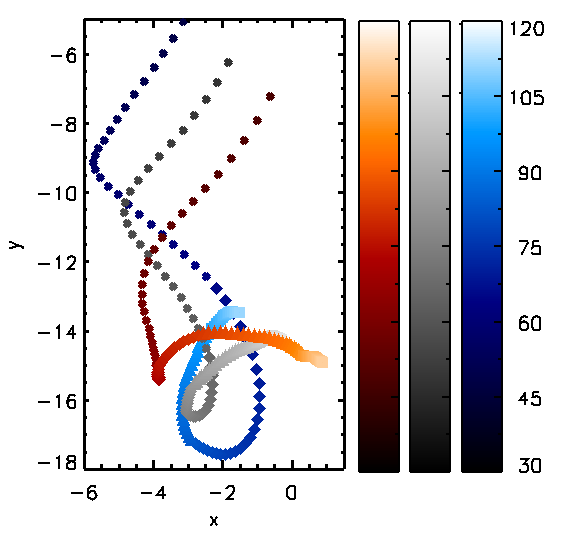}
\subcaption{Fieldline trajectories.}
\label{fig:rotfieldtrajectorya}
\end{subfigure}\hspace{-0.2cm}~
\begin{subfigure}{0.22\textwidth}
\centering
\includegraphics[scale=0.23]{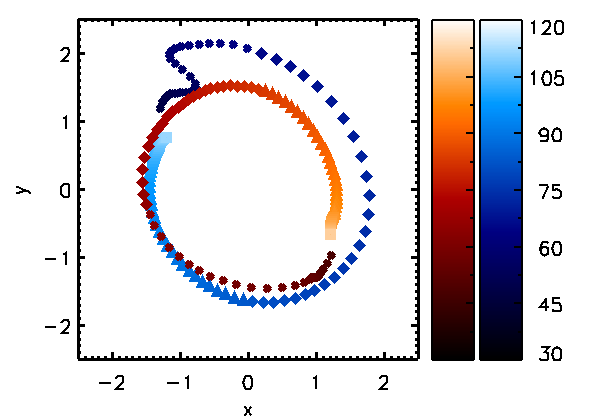}
\subcaption{Relative fieldline trajectories with black axis location subtracted.}
\label{fig:rotfieldtrajectoryb}
\end{subfigure}
\caption{The trajectories of the fieldlines as they pass through the photospheric plane coloured with increasing intensity as time progresses. The colour scale on the right shows the times during the evolution.}
\label{fig:rotfieldtrajectories}
\end{figure}
These figures clearly demonstrate the rotation of the fieldlines threading the sunspots in a visual manner. Examining Figure~\ref{fig:rotsunspot}, it appears that both the red and blue fieldlines have rotated through an angle of at least $\pi$ radians over $40$ time steps. 

In order to follow the fieldlines undergoing this rotation, we have traced the $x$ and $y$ coordinates of the locations of the red, black and blue fieldlines as they pass through the photospheric plane. The trajectories of these fieldlines are shown in Figure~\ref{fig:rotfieldtrajectorya}. Initially, the three fieldlines drift outwards in a line as the sunspots separate. Later, the motions slow down and rotation develops. This is not particularly clear given the translational aspect of the motion as the sunspots separate. To remove this feature of the motion, we have subtracted off the location of the black axis from the position of the blue and red fieldlines in Figure~\ref{fig:rotfieldtrajectoryb}. This gives us the relative position of the blue and red fieldlines and indicates that the blue and red fieldlines rotate clockwise around the black axis by an angle of almost $360\degree$.

\begin{figure}[ht]
\centering
\includegraphics[width=.15\textwidth]{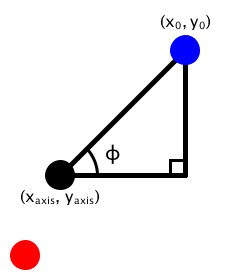}
\caption{Representation of angle $\phi$.}
\label{fig:rottanphi}
\end{figure}

\begin{figure}[ht]
\centering
\includegraphics[width=8cm]{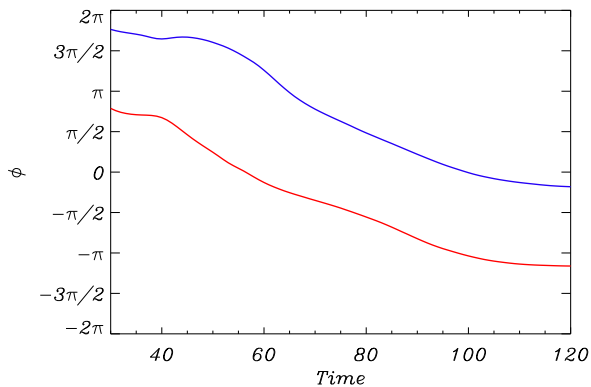}
\caption{Evolution of angle of rotation $\phi$ for both the red and blue fieldlines as depicted in Figures~\ref{fig:rotsunspotsfielda} and~\ref{fig:rotsunspotsfieldb}.}
\label{fig:rotangle}
\end{figure}

With the $x$ and $y$ coordinates of the intersections of selected fieldlines through the photosphere, we can calculate the angle of rotation using
\begin{equation}
\tan{\phi} = \frac{y_{0}-y_{\text{axis}}}{x_0-x_{\text{axis}}},
\label{eq:rottanphi}
\end{equation}
where $x_{\text{axis}}$ and $y_{\text{axis}}$ are the $x$ and $y$ coordinates of the axis of the tube (black fieldline) and $x_{0}$ and $y_{0}$ are the coordinates of the fieldline we are investigating, i.e. the red or blue fieldline. In Figure~\ref{fig:rottanphi} a schematic has been included to help us visualise the meaning of the angle $\phi$. Through the use of equation~\eqref{eq:rottanphi}, the angle $\phi$ has been calculated for both the red and blue fieldline as displayed in Figure~\ref{fig:rotangle}.
As the two chosen fieldlines were initially equally spaced on either side of the axis, the rotation angles are $\pi$ out of phase at the beginning. Both fieldlines undergo a rotation of between $7\pi/4$ and $2\pi$ over $90$ time steps. More precisely, the red fieldline undergoes a rotation of $340\degree$ and the blue fieldline undergoes a rotation of $353\degree$. 

\begin{figure}[ht]
\centering
\includegraphics[width=8cm]{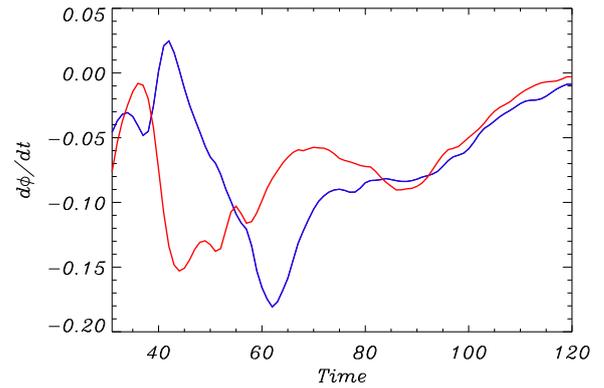}
\caption{Evolution of rate of change of the angle of rotation $\phi$ for both the red and blue fieldlines as shown in Figures~\ref{fig:rotsunspotsfielda} and~\ref{fig:rotsunspotsfieldb}.}
\label{fig:rotrates}
\end{figure}

Next, we consider the temporal rate of change of the angle of rotation, i.e. $\mathrm{d}\phi/\mathrm{d}t$, as shown in Figure~\ref{fig:rotrates}.
This illustrates that different regions of the sunspot are rotating at slightly different rates. There is an initial peak in the size of the rotation rate of both fieldlines of between $-0.15$ and $-0.2$. This peak in rotation rate occurs at about $t=44$ for the red fieldline and at about $t=62$ for the blue fieldline. The rate of rotation diminishes as the experiment proceeds until it reaches zero indicating that the fieldlines have essentially stopped rotating. 

\begin{figure}[ht]
\centering
\begin{subfigure}{0.24\textwidth}
\centering
\includegraphics[width=4.8cm]{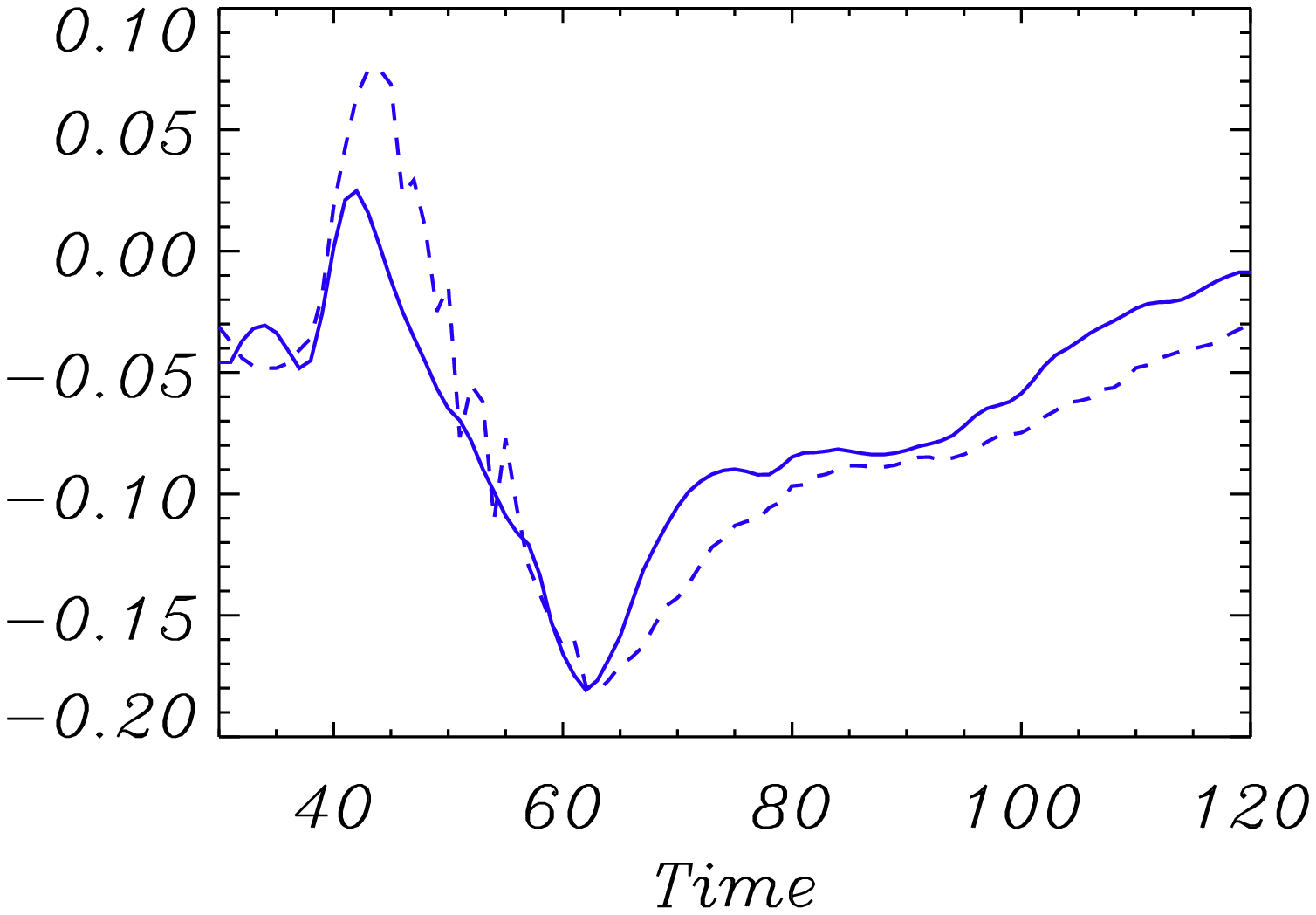}
\subcaption{Blue fieldline traced from ($0,-14.5,-25)$.}
\label{subfig:rotvortzred}
\end{subfigure}~
\begin{subfigure}{0.24\textwidth}
\centering
\includegraphics[width=4.8cm]{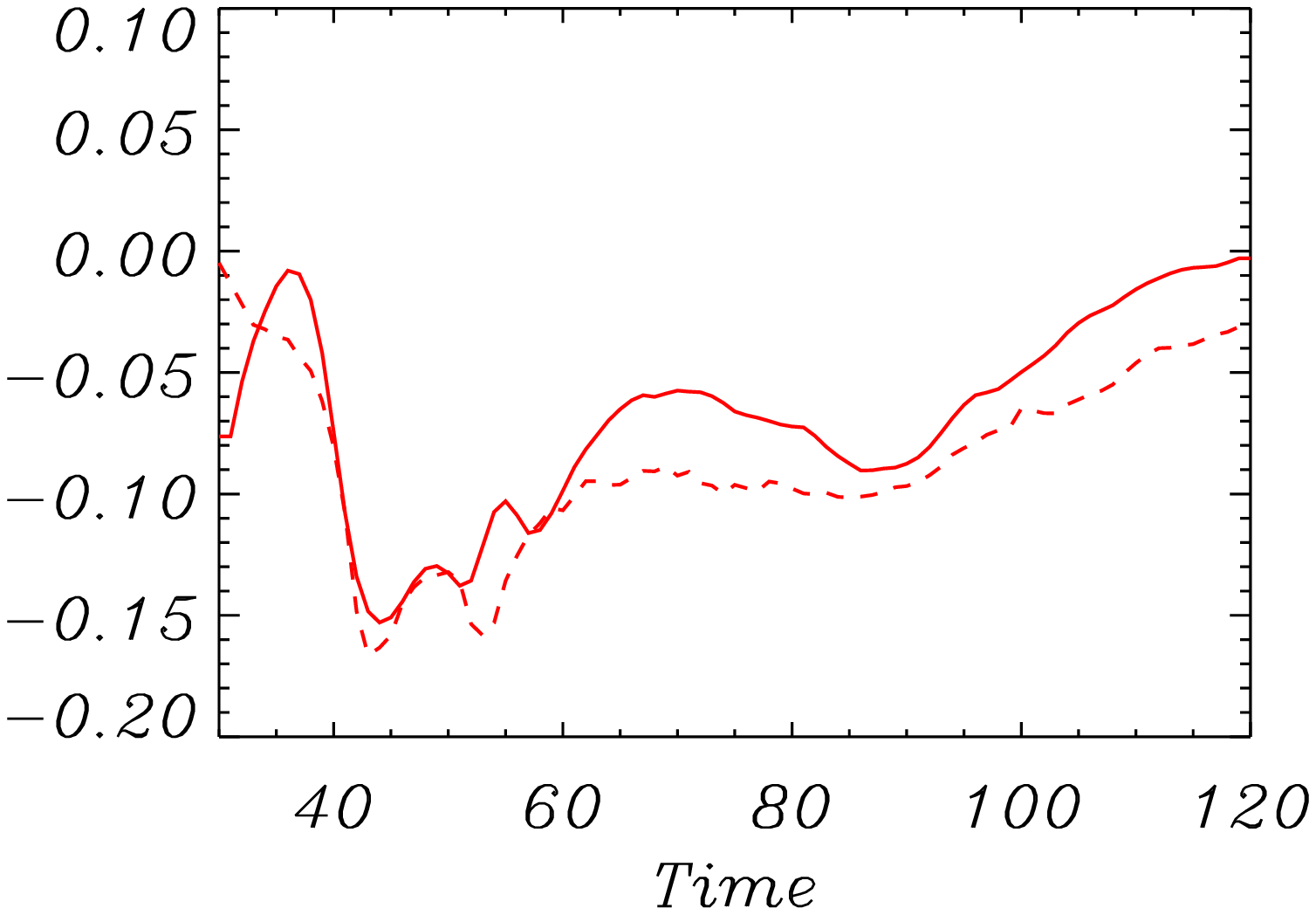}
\subcaption{Red fieldline traced from ($0,-15.5,-25)$.}
\label{subfig:rotvortzblue}
\end{subfigure}
\caption{Comparison of terms $\displaystyle{\frac{\mathrm{d}\phi}{\mathrm{d}t}}$ (solid line) and $\displaystyle{\frac{\omega_z}{2}}$ (dashed line) for both the blue and red fieldlines respectively.}
\label{fig:rotvortzrate}
\end{figure}
This rotation is investigated further by checking if the assumption of solid body rotation is reasonable, i.e. that the rotation angle does not depend on the radius of the fieldline. We have concluded from above that different areas of the sunspot are rotating at slightly different rates. If we assume the rotation is a solid body rotation then it follows that the velocity in the $\phi$ direction, at radius $R$, is given by
\begin{equation}
v_{\phi} = R\frac{\mathrm{d}\phi}{\mathrm{d}t},\nonumber
\end{equation}
and the $z-$component of the vorticity is given by
\begin{eqnarray}
\omega_z & = & \frac{1}{R}\frac{\partial }{\partial R}\left(Rv_{\phi}\right) = 2\frac{\mathrm{d}\phi}{\mathrm{d}t},\nonumber
\end{eqnarray}
where we have assumed that $\phi$ does not depend on $R$. We can therefore relate the vertical vorticity and the rate of change of the angle $\phi$ by
\begin{equation}
\frac{\mathrm{d}\phi}{\mathrm{d}t} = \frac{\omega_z}{2}.
\label{eq:rotrates}
\end{equation}
To check if our assumption is valid, we can investigate equation~\eqref{eq:rotrates} by plotting $\mathrm{d}\phi/\mathrm{d}t$ and $\omega_z/2$ for both the red and blue fieldlines. 
In both panels of Figure~\ref{fig:rotvortzrate}, the two terms balance each other well suggesting that equation~\eqref{eq:rotrates} is valid and the rotation angle may not have a large dependence on the radius from the axis of the tube.
\subsection{Vorticity}
In another attempt to demonstrate the rotational movement at the photosphere, we analyse the plasma motion on the photospheric plane. Hence, the vorticity, calculated as the curl of the plasma velocity, is examined as this quantifies the rotation of the plasma. As we are concerned with the rotation of sunspots, the vertical component of the vorticity is the quantity of interest, as it measures the rotation in the $x-y$ plane. This is expressed as,
\begin{equation}
\omega_z=(\mathbf{\nabla} \times \mathbf{v})_{z} = \frac{\partial v_y}{\partial x} - \frac{\partial v_x}{\partial y}.\nonumber
\label{eq:vortz}
\end{equation}
A positive $\omega_z$ refers to a counter-clockwise motion while a negative $\omega_z$ refers to a clockwise motion. In our experiment, the initial field is right-hand twisted; that is, if we were to create this field from a straight field, both footpoints would have been rotated in a clockwise motion. In other words, the fieldlines are wound counter-clockwise in the positive $B_z$ footpoint and clockwise in the negative $B_z$ footpoint. 

In order to visualise the field, let us consider the schematic of the fieldlines soon after the field has intersected the photosphere as shown in Figure~\ref{fig:rotinteriorfieldt40}. At this point, the legs of the tube have started to straighten out and almost resemble a vertical cylindrical tube originating at $z=-25$ and intersecting the photosphere at $z=0$. As the magnetic field is fixed at the base of the computational domain, a clockwise rotation at the photosphere is required in order to unwind the interior portion of the field. The atmospheric field, on the other hand, is twisted up by a clockwise rotational motion on the two footpoints at the photosphere.

\begin{figure}[ht]
\centering
\begin{subfigure}{0.24\textwidth}
\centering
\includegraphics[width=4.5cm]{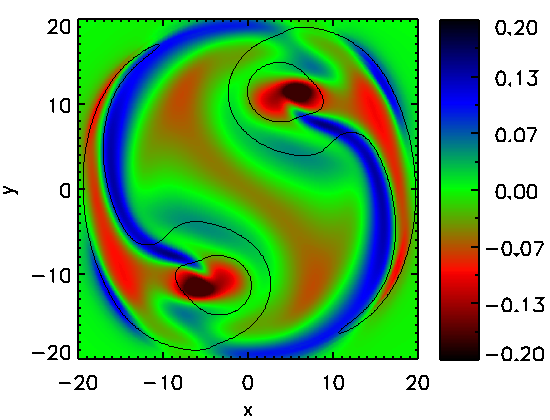}
\caption{$t=40$}
\end{subfigure}~
\begin{subfigure}{0.24\textwidth}
\centering
\includegraphics[width=4.5cm]{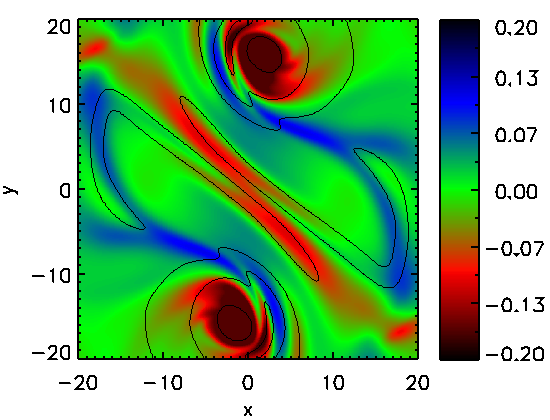}
\caption{$t=80$}
\end{subfigure}~
\caption{Coloured contours of $\omega_z$ accompanied by line contours of $B_z$ at $z=0$, the base of the photosphere, for the specified times. A movie of this figure is included in the electronic version.}
\label{fig:vortz}
\end{figure}

A series of coloured contour plots of the vertical vorticity are displayed in Figure~\ref{fig:vortz} at times $t=40$ and $t=80$. For visualisation purposes, line contours of $B_z$ have been over-plotted to show the location of the sunspots. At the centre of each of the polarities, there is a concentration of negative $\omega_z$ corresponding to a clockwise rotation. This concentration of $\omega_z$ appears when the legs of the tube straighten and builds with time until it peaks at about $t=50$ before decaying as the simulation progresses. This hints that there is some bulk rotation of the sunspots, similar to the sunspot rotations observed by~\cite{Brown2003} and~\cite{Yan2009}. This result complements our previous investigation of the rotation angle where we found fieldlines undergoing significant rotations. Interestingly the same sign of vertical vorticity is present in both concentrations of opposite polarity. As discussed earlier, this has important consequences for the evolution of the field in both the atmosphere and interior. We suggest that these sunspot rotations are due to the untwisting of the field in the interior injecting twist into the atmosphere. Another interesting feature of the contour plots is the red streak of negative vorticity between the sunspots. Elongated streaks of vorticity most likely correspond to shearing motions rather than rotation suggesting that these are due to shear flows between the sunspots. In addition, there are blue tails of positive vorticity located on the inner side of each of the sunspots again indicative of shearing motions. 

\begin{figure}[ht]
\centering
\includegraphics[scale=0.5]{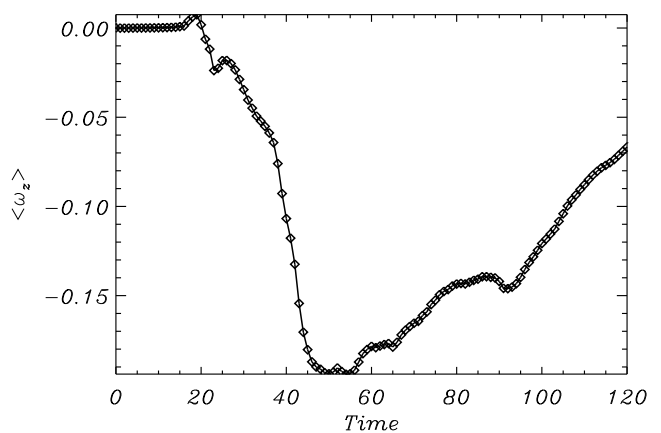}
\caption{Evolution of the mean vertical vorticity $\langle \omega_z \rangle$ averaged over the area of the upper polarity concentration where $B_z$ is above $75\%$ of the max of $B_z$ on $z=0$ as described in equation~\eqref{eq:averagevortz}.}
\label{fig:rotavevortz0}
\end{figure}

Now that we have established a clear rotational velocity at the photosphere, the evolution of the vertical vorticity at the photosphere is expressed in a more quantifiable manner. Following the method used by~\cite{Fan2009}, we achieve this by plotting the time variation of the mean vertical vorticity, $\langle\omega_z\rangle$, averaged over the area of the upper sunspot where $B_z$ is greater than $3/4$ of its peak value in Figure~\ref{fig:rotavevortz0}. Explicitly,
\begin{equation}
\langle\omega_z\rangle = \frac{1}{N}\left(\sum_{k=1}^{N}{\omega_z(x_{k},y_{k},z=0)}\right),
\label{eq:averagevortz}
\end{equation}
where $x_{k}$ and $y_{k}$ are the $x$ and $y$ coordinates of the region satisfying $B_z > \frac{3}{4}\text{max}(B_z)$ and $N$ is the number of points that satisfy this criteria. This has been compared to the lower sunspot with no notable difference in result.

Considering the average vertical vorticity at the photosphere in Figure~\ref{fig:rotavevortz0}, the vorticity is consistently negative suggesting that the dominant motion is a clockwise rotation. We find that a clockwise vortical motion appears in each polarity source when the field reaches the photosphere. The vortical motion quickly rises to a peak at roughly $t=50$ soon after the emerged field becomes vertical and the photospheric footprints have reached their maximum separation. At this time rapid rotation commences. The horizontal velocity at this time is approximately $0.5$ in magnitude which corresponds to a physical velocity of $3.4$~km~$\text{s}^{-1}$. Soon after, the vorticity steadily begins to decline. This significant clockwise rotation twists up the emerged fieldlines in the atmosphere transporting twist from the tube's interior portion to its stretched coronal portion. As noted earlier, we speculate that this transport of twist is due to some form of torsional Alfv\'en wave. 

 As discussed earlier,~\cite{Min2009} conjectured that the observed rotation of sunspots due to flux emergence may be an apparent effect due to a twisted field rising and each fieldline appearing in a different position at the photosphere. However, in our experiment we have established a clear rotation in the plasma velocity suggesting that this may not be the case. To estimate the contribution to the rotation by apparent effects, we quantify the vertical advection of the flux tube by averaging the vertical velocity in a similar fashion to the way in which we averaged the vorticity. To achieve an upper bound for our estimate, we take the vertical speed of the tube to be the maximum of $\langle v_z \rangle$ and assume that the vertical leg has a full turn of twist at $t=40$ when the field intersects. This is equivalent to the field being advected vertically by $2.4$ units by the end of the experiment, corresponding to a $34.6\degree$ apparent rotation. This is an over-estimate for the apparent rotation angle as we have assumed the maximum velocity for all time, and yet, this is still significantly smaller than the calculated rotation angle. To conclude, this helps us disregard this theory and explain the rotation solely by a dynamical consequence of the emergence of flux. 

Though the vorticity plots do give us a useful insight into the rotational properties of the plasma at the photosphere, further examination is necessary in order to quantify the untwisting of the interior field, and the transport of twist into the atmosphere.

\subsection{Current density}
\begin{figure}[ht]
 \centering
 \begin{subfigure}{0.24\textwidth}
 \centering
 \includegraphics[width=4.5cm]{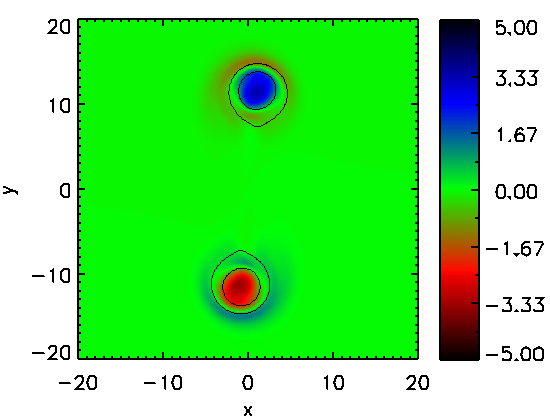}
 \subcaption{$z=-12.5$ and $t=40$}
 \label{subfig:rotjz-1240}
 \end{subfigure}~
 \begin{subfigure}{0.24\textwidth}
 \centering
 \includegraphics[width=4.5cm]{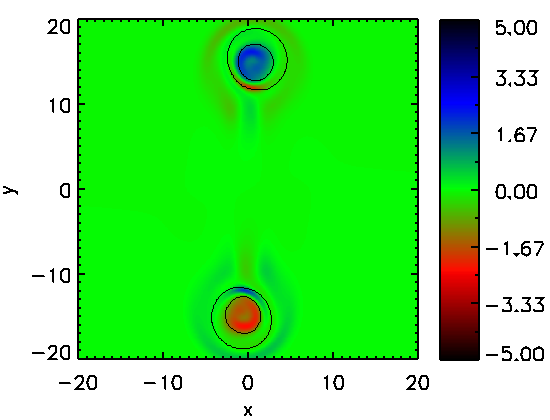}
 \caption{$z=-12.5$ and $t=80$}
 \label{subfig:rotjz-1280}
 \end{subfigure}\\
 \begin{subfigure}{0.24\textwidth}
 \centering
 \includegraphics[width=4.5cm]{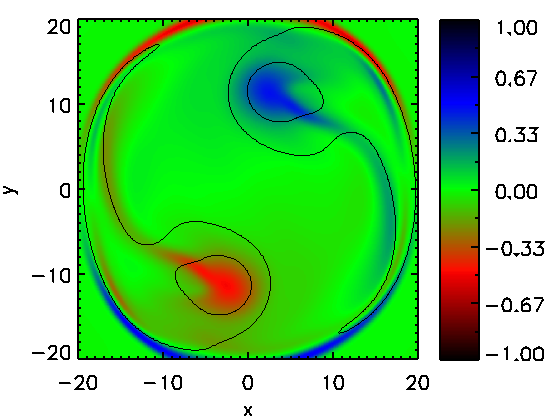}
 \subcaption{$z=0$ and $t=40$}
 \label{subfig:rotjz040}
 \end{subfigure}~
 \begin{subfigure}{0.24\textwidth}
 \centering
 \includegraphics[width=4.5cm]{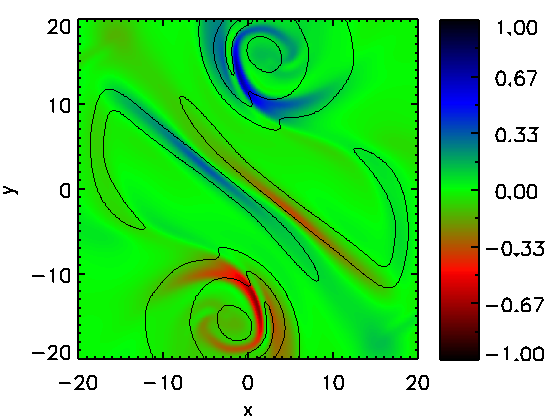}
 \caption{$z=0$ and $t=80$}
 \label{subfig:rotjz080}
 \end{subfigure}
 \caption{Coloured contours of $j_z$ at the plane in the middle of the interior ($z=-12.5$) in the top panel and at the solar surface ($z=0$) in the bottom panel for the specified times, as well as line contours of $B_z$ for comparison of the size of sunspots.}
 \label{fig:rotjz}
 \end{figure}

A useful quantity when estimating the twist of the magnetic field is the current density, specifically the  $z$-component, as denoted by
 \begin{equation}
 j_z = \frac{1}{\mu}\left(\frac{\partial B_y }{\partial x} - \frac{\partial B_x}{\partial y}\right),\nonumber
 \end{equation}
 as this is related to how twisted the field is in the $x-y$ plane. We consider coloured contours of $j_z$ at a height half way down the solar interior and at the base of the photosphere, as shown in Figure~\ref{fig:rotjz}.
 From the coloured contours of $j_z$ in Figure~\ref{fig:rotjz}, it is clear that although the majority of each sunspot is positive or negative, the periphery of each sunspot is dominated by the opposite sign of $j_z$. As we insist that our initial sub-photospheric flux tube is isolated and therefore surrounded by unmagnetised plasma, Faraday's law demands that the flux tube must carry no net current. As the flux tube carries current inside due to its twist, a reverse current surrounds the sunspot to ensure a zero net current in this region. Focussing on the top panel of Figure~\ref{fig:rotjz}, there is some evidence that the two concentrations of strong $j_z$ centred on the sunspots are depleting with time. Similarly, considering the bottom panel of Figure~\ref{fig:rotjz}, the concentrations of $j_z$ at the photosphere intensify when the field first emerges then diminish as the experiment proceeds. As $j_z$ signifies the twist of the field in the $x-y$ direction and is related to the azimuthal magnetic field, a decrease in $j_z$ could indicate a decline in the amount of twist. The mechanism responsible for this decrease in $j_z$ requires further investigation.
 
 \begin{figure}[ht]
 \centering
 \includegraphics[scale=0.5]{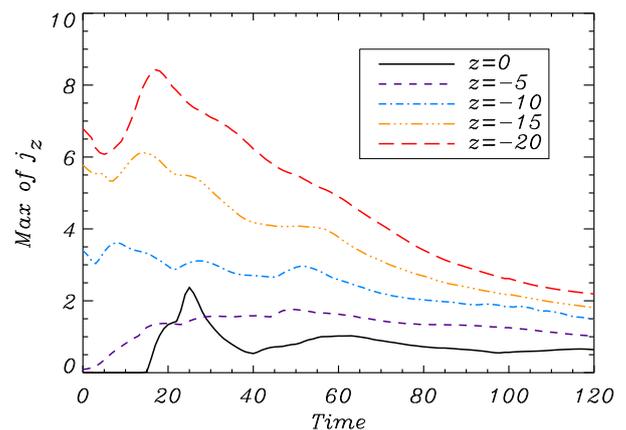}
 \caption{Plot of the maximum value of $j_z$ against time for varying heights depicted in the key. }
 \label{fig:maxjzt}
 \end{figure}
 
\begin{figure}[ht]
 \centering
 \includegraphics[scale=0.5]{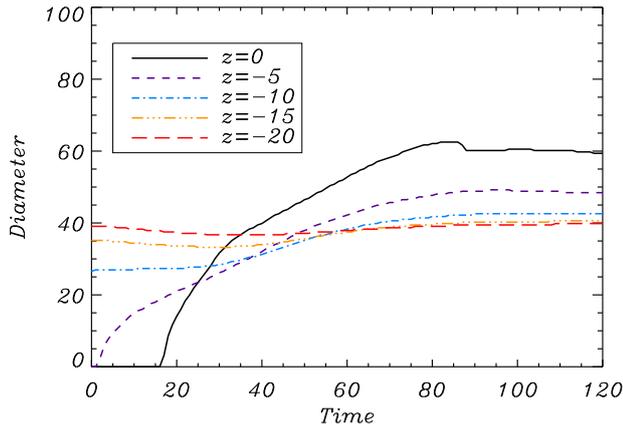}
 \caption{Line plot of the diameter of the {\bf $B_z=1.0$} contour as a function of time for varying heights depicted in the key.}
 \label{fig:rotdiambz}
 \end{figure}

The temporal variation of the maximum value of $j_z$ for specific heights below the photosphere is displayed in Figure~\ref{fig:maxjzt}. There is an initial increase in the maximum of $j_z$ for all heights due to the emergence of the field before a steady decline as the experiment proceeds. There are two possible explanations for this steady decrease in the vertical current. This could be caused by the expansion and stretching of the field as a result of emergence or by a decline in the amount of twist stored in the field. The stretching of the field results in a decline in the gradients of $B_x$ and $B_y$ and lowers $j_z$. To evaluate the extent of the expansion of the field, we estimate the diameter of a contour of $B_z$ as shown in Figure~\ref{fig:rotdiambz}. This plot shows the maximum y-separation of the contour of $B_z=1$. Initially, the separation increases for heights near the photosphere as the flux tube buoyantly rises and the legs of the tube straighten. Later, there is very little change in the separation of the legs for heights deep in the solar interior. There is however some expansion of the magnetic field for the photospheric height $z=0$ and $5$ units below the boundary as expected as the magnetic field expands into the low density atmosphere. This helps us disregard this cause and explain the decrease in $j_z$ in the solar interior solely by an untwisting of the field. Specifically, a decrease in $j_z$ implies a decline in the azimuthal field which corresponds to the interior portion of the field untwisting. However, a decrease in $j_z$ at the photosphere may be explained by the expansion of the field in this region.

 \begin{figure}[ht]
 \centering
 \begin{subfigure}{0.48\textwidth}
 \centering
 \includegraphics[width=7cm]{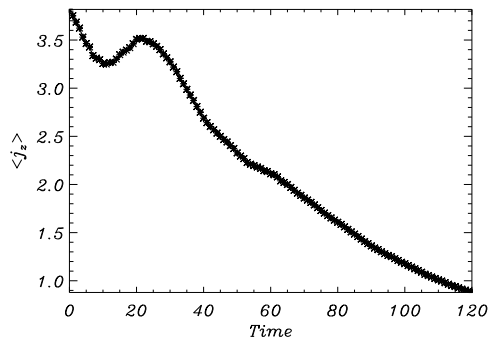}
 \subcaption{$z=-12.5$}
 \end{subfigure}\\
 \begin{subfigure}{0.48\textwidth}
 \centering
 \includegraphics[width=7cm]{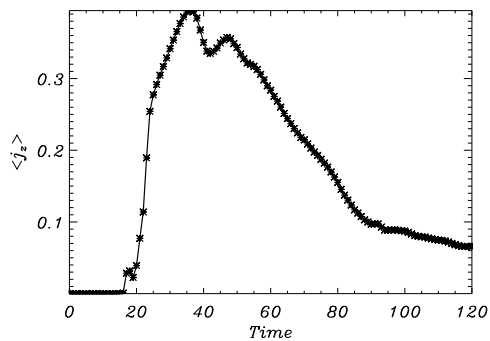}
 \caption{$z=0$}
 \end{subfigure}
 \caption{Time evolution of the vertical current $\langle{j_z}\rangle$ averaged over the area of the positive polarity flux source where $B_z$ is greater than $75\%$ of its maximum.}
 \label{fig:rotaveragejz}
 \end{figure}
 
 Finally, to analyse the current further, an examination of the total $j_z$ in each sunspot is necessary. As we noted from the coloured contours, although the centre of each sunspot is dominated by one sign of current, the outer boundary consists of reverse current. Hence, we cannot estimate the current in each sunspot by measuring the total positive or negative current. Instead, we estimate the current in the centre of the upper sunspot by averaging the vertical current over the area where the vertical magnetic field is greater than $3/4$ of its maximum in a similar fashion to our calculation of the average vorticity. 
The temporal evolution of this quantity is depicted in Figure~\ref{fig:rotaveragejz} for the interior plane ($z=-12.5$) and the photospheric plane respectively.
 
 At the photosphere, the mean vertical current increases as the magnetic field emerges then steadily declines as the field expands into the atmosphere. Lower down in the solar interior, the mean vertical current generally declines in a linear manner. After an initial drop in $\langle{j_z}\rangle$, there is a small increase as the field straightens out and the negative outer-boundary plays a less dominant role. Overall, there is a general decrease in $\langle{j_z}\rangle$ suggesting a drop in the twist stored in the interior portion of the field as the field untwists.

\subsection{Magnetic helicity}

\label{sec:hel}

In order to quantify the transport of twist from the solar interior to the solar atmosphere due to the emergence and untwisting of the field in the interior, we have calculated the evolution of the relative magnetic helicity both above and below the photosphere. The magnetic helicity is a topological quantity describing how much a magnetic structure is twisted, sheared or braided. 

The relative magnetic helicity (to the reference field $\mathbf{B}_p$) of the field $\mathbf{B}$ in a given volume $V$ is given by~(\cite{Berger1984},~\cite{Finn1985})
\begin{equation}
H_{r}= \int_{V}{(\mathbf{A}+\mathbf{A}_{p}) \cdot (\mathbf{B} - \mathbf{B}_p)\text{ }\mathrm{d}V},
\label{eq:rotrelhelicity}
\end{equation}
where $\mathbf{A}$ is the vector potential of $\mathbf{B}$ ($\mathbf{B} = \bf{\nabla} \times \mathbf{A}$), $\mathbf{B}_p$ is the reference potential field with the same normal flux distribution as $\mathbf{B}$ on all bounding surfaces and $\mathbf{A}_{p}$ is the vector potential of $\mathbf{B}_p$ ($\mathbf{B}_p = \mathbf{\nabla} \times \mathbf{A}_p$). The relative magnetic helicity is favoured over the magnetic helicity as it has been shown to be gauge-independent with respect to the gauges for $\mathbf{A}$ and $\mathbf{A}_{p}$. This is a necessary condition for a physically meaningful definition of helicity.

In order to calculate the relative magnetic helicity numerically, we have tested and compared two approaches; the calculation of $\mathbf{A}$ and $\mathbf{A}_p$ using DeVore's method~\citep{DeVore2000} and the numerical procedure from~\cite{Moraitis2014}. As the results were comparatively similar, we will discuss the calculation of the relative magnetic helicity from~\cite{Moraitis2014}. 

In calculating the potential field within the volume $V=[x_1,x_2]~\times~[y_1,y_2]~\times~[z_1,z_2]$, the numerical procedure utilised in~\cite{Moraitis2014} takes into account all boundaries within the finite volume. This is advantageous over DeVore's method which is only valid for a semi-infinite space above a lower boundary. The potential field satisfies $\mathbf{j}_p = \mathbf{\nabla} \times \mathbf{B}_p = \mathbf{0}$ within $V$, thus implying $\mathbf{B}_p = -\mathbf{\nabla} \phi$ where $\phi$ is a scalar potential. The solenoidal constraint $\nabla \cdot \mathbf{B}_p =0$ then implies that the scalar potential is a solution of Laplace's equation $\nabla^2 \phi =0$ in $V$. The condition that $\mathbf{B}$ and $\mathbf{B}_p$ have the same normal components along the boundaries of the volume translates to Neumann boundary conditions for $\phi$, i.e. $ \left.\partial \phi/\partial \hat{n}\right|_{\partial V} = -  \left. \hat{n} \cdot \mathbf{B}\right|_{\partial V}$. The Laplace equation is then solved numerically using a standard FORTRAN routine included in the FISHPACK library~\citep{Swartztrauber1979}.

The original and potential fields are now stored for the given time step and desired volume. The next step is to calculate the corresponding vector potentials given the method proposed by~\cite{Valori2012}. As the relative magnetic helicity is gauge-independent, we are free to choose the gauge $\mathbf{A} \cdot \hat{z} = 0$ throughout $V$ so that the $x$ and $y$ components of $\mathbf{B} = \mathbf{\nabla} \times \mathbf{A}$ are integrated over the interval $(z_1,z)$ to 
\begin{equation}
\mathbf{A} = \mathbf{A}_0 - \hat{z}~\times\int_{z_1}^{z}{\mathbf{B}(x,y,z')~\mathrm{d}z'},\nonumber
\end{equation}
where $\mathbf{A}_0 = \mathbf{A}(x,y,z=z_1) = (A_{0x},A_{0y},0)$ is a solution to the $z$-component of $\mathbf{B} = \mathbf{\nabla} \times \mathbf{A}$, i.e. 
\begin{equation}
\frac{\partial A_{0y}}{\partial x} - \frac{\partial A_{0x}}{\partial y}=B_z(x,y,z=z_1).\nonumber
\end{equation}
Following Valori's method we choose the simplest solution to the above equation, given by
\begin{eqnarray*}
A_{0x}(x,y,z=z_1)& = &-\frac{1}{2}\int_{y_1}^{y}B_z(x,y',z=z_1)~\mathrm{d}y',\\
A_{0y}(x,y,z=z_1) &= &\frac{1}{2}\int_{x_1}^{x}B_z(x',y,z=z_1)~\mathrm{d}x'.
\end{eqnarray*}
Similarly, the vector potential of the potential field is calculated using
\begin{equation}
\mathbf{A}_p = \mathbf{A}_0 - \hat{z}~\times\int_{z_1}^{z}{\mathbf{B}_p(x,y,z')~\mathrm{d}z'},\nonumber
\end{equation}
where we have noted that $\mathbf{A}_{p0}=\mathbf{A}_0$ as $\mathbf{B}$ and $\mathbf{B}_p$ share the same normal component on the boundary at $z=z_1$.

An alternative solution for the vector potentials can be obtained if we use the top boundary, i.e. integrating over the interval $(z,z_2)$ as
\begin{equation}
\mathbf{A} = {\text{\bf\~{A}}}_{0}+ \hat{z}~\times \int_{z}^{z_2}{\mathbf{B}(x,y,z')~\mathrm{d}z'}.\nonumber
\end{equation}
This has been checked for comparison and there is no notable difference between the two solutions.

\begin{figure}[ht]
\centering
\includegraphics[scale=0.4]{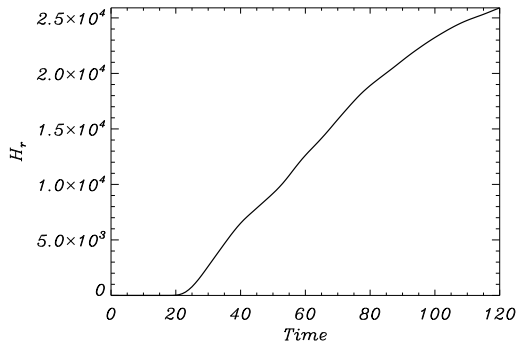}
\caption{Evolution of the relative helicity $H_r$ when calculated above $z=0$ in the solar atmosphere.}
\label{fig:rothelicityatmosphere}
\end{figure} 

With the corresponding potential and vector potentials for the magnetic field, we can calculate the relative magnetic helicity within different subvolumes of the total simulation domain. First, we calculate the atmospheric magnetic helicity above the photosphere, as shown in Figure~\ref{fig:rothelicityatmosphere}. As expected, the atmospheric helicity is zero until $t=20$ when the magnetic field first emerges. Subsequently, there is a linear increase in helicity as twist is steadily injected into the atmosphere by the emergence of flux and rotational motions at the photosphere. By the end of the experiment, the magnetic helicity injected into the atmosphere has reached $3.6\times10^{23}$ Wb$^2$ ($3.6\times10^{39}$ Mx$^2$), typical of a small event. This can be compared with observations where~\cite{Min2009} quote a helicity transport of $4\times10^{42}$ Mx$^2$ when considering a much larger active region. To investigate the normalised magnetic helicity that is transported to the atmosphere, we divide by $\Phi_{\text{tube}}^2$ and find that the evolution follows the same linear monotonic shape and reaches a value of $0.83$ by $t=120$ corresponding to almost one full twist of the flux in the original tube.

\begin{figure}[ht]
\centering
\includegraphics[scale=0.4]{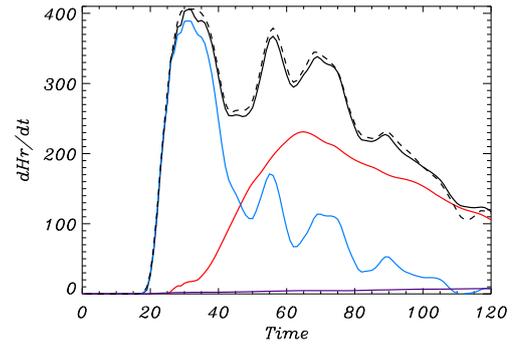}
\caption{Evolution of the time derivative of the relative helicity $H_r$ when calculated above $z=0$ in the solar atmosphere using equation~\eqref{eq:rotdhdtf}. The total time derivative (black solid line) is split into the \emph{dissipation} term (purple solid line), the \emph{surface correction} term (yellow solid line), the \emph{shear} term (red solid line) and \emph{emergence} term (blue solid line). The dashed black line is the derivative of the curve from Figure~\ref{fig:rothelicityatmosphere} calculated using finite differencing.}
\label{fig:rotdhdtatmospheref}
\end{figure}

To further investigate the magnetic helicity in the atmosphere we have calculated the time derivative of the magnetic helicity above the photosphere, both numerically using finite differencing and analytically. This helps us to understand the main sources contributing to the production and depletion of helicity. Magnetic helicity is mainly contributed to by vertical flows that advect twisted magnetic flux into the corona and by surface flows that shear and twist magnetic fields~\citep{Berger1984}. The rate of change of relative magnetic helicity, $H_r$, can be evaluated analytically by differentiating the expression in~\eqref{eq:rotrelhelicity}~\citep{Berger1984},
{\small{
\begin{eqnarray}
\frac{\mathrm{d}H_{\text{r}}}{\mathrm{d}t} &=& - 2\eta\int{\mathbf{j} \cdot \mathbf{B}~ \mathrm{ d}V} + 2\eta\int{((\mathbf{A}_{p} \times \mathbf{j}) \cdot \mathbf{n}~\mathrm{ d}S}\nonumber\\
& + & 2\int{\left[(\mathbf{A}_p \cdot \mathbf{v})(\mathbf{B} \cdot \mathbf{n}) -  ((\mathbf{A}_p \cdot \mathbf{B})(\mathbf{v} \cdot \mathbf{n})\right]~ \mathrm{d }S}.
\label{eq:rotdhdtf}
\end{eqnarray}}}
The first term on the right-hand side of equation~\eqref{eq:rotdhdtf} relates to the depletion of helicity by internal dissipation (\emph{dissipation} term), the second corresponds to a surface correction to the resistive dissipation (\emph{surface correction} term) , the third relates to the generation of helicity by horizontal motions of the boundary (\emph{shear} term) and the last corresponds to the injection of helicity by direct emergence (\emph{emergence} term). Let us consider the rate of change of atmospheric helicity in Figure~\ref{fig:rotdhdtatmospheref}. From the point the field emerges until $t=45$, the helicity flux due to emergence (blue solid line) dominates the change in helicity. Later, the horizontal shearing and rotational motions at the photospheric footpoints (red solid line) are the primary sources of helicity change, in agreement with~\citep{Fan2009}. The change in helicity due to internal helicity dissipation (purple) and the surface correction term (yellow) are much less significant and do not contribute to the overall change in helicity in the atmosphere. The derivative of $H_r$ from Figure~\ref{fig:rothelicityatmosphere} is over plotted for comparison. The two approaches agree very well indicating that numerical effects are kept to a minimum. Care must be taken when considering the rate of change of helicity given in equation~\eqref{eq:rotdhdtf}. See~\cite{Pariat2015} for the full derivative including additional terms. Clearly, from Figure~\ref{eq:rotdhdtf}, the additional terms are not important in this particular case as the flux through the surfaces closely follows the time derivative of the helicity. ~\cite{Pariat2015} also notes that precaution must also be taken when dividing the helicity flux into individual terms as although their sum is gauge-independent the individual terms are not, hence, limiting their physical meaning.

\begin{figure}[ht]
\centering
\begin{subfigure}{0.48\textwidth}
\centering
\includegraphics[scale=0.38]{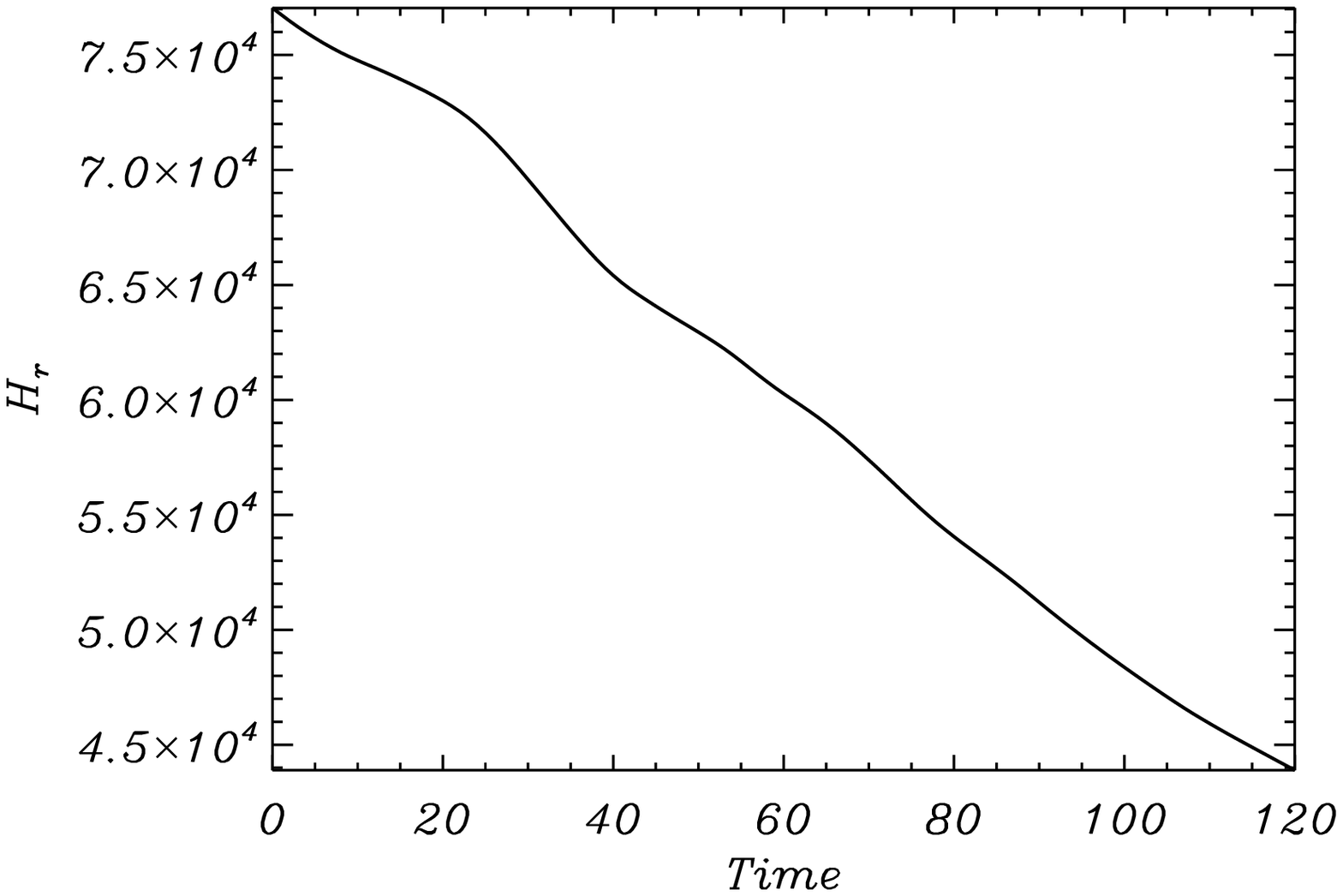}
\subcaption{$H_r$ }
\end{subfigure}\\
\begin{subfigure}{0.48\textwidth}
\centering
\includegraphics[scale=0.38]{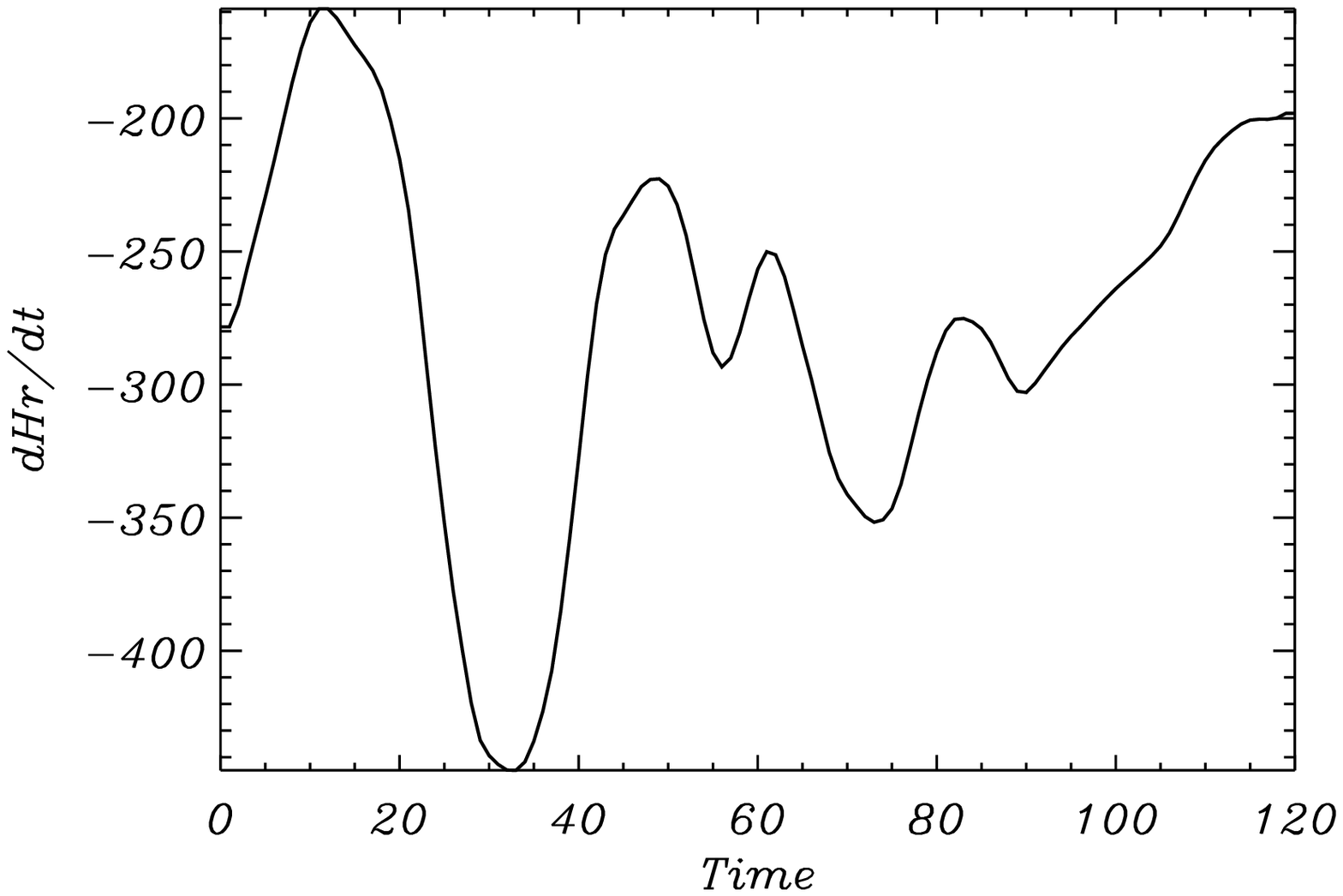}
\subcaption{$\displaystyle{\mathrm{d}H_r/\mathrm{d}t}$ }
\end{subfigure}
\caption{Evolution of the relative helicity and rate of change of helicity when calculated below $z=0$ in the solar interior.}
\label{fig:rothelicityinterior}
\end{figure}

Now that we have established a clear increase in the helicity in the atmosphere, we analyse the helicity in the solar interior. As the toroidal flux tube has shown clear signs of untwisting, we expect that the magnetic helicity in the interior will decrease due to both the emergence of magnetic flux and the rotational movements on the photospheric boundary. The calculation of the relative magnetic helicity and the change in helicity below the photosphere are shown in Figure~\ref{fig:rothelicityinterior}.
As expected, the interior relative magnetic helicity decreases throughout the experiment. Initially, the decrease in magnetic helicity is dominated by internal helicity dissipation. Later, at $t=20$, the field emerges from the interior through to the atmosphere resulting in a sharper decrease due to both the emergence of magnetic flux and the rotational movements on the photosphere extracting helicity from the interior as the flux tube untwists.

\subsection{Magnetic energy}
\begin{figure}[!h]
\centering
\begin{subfigure}{0.48\textwidth}
\centering
\includegraphics[scale=0.38]{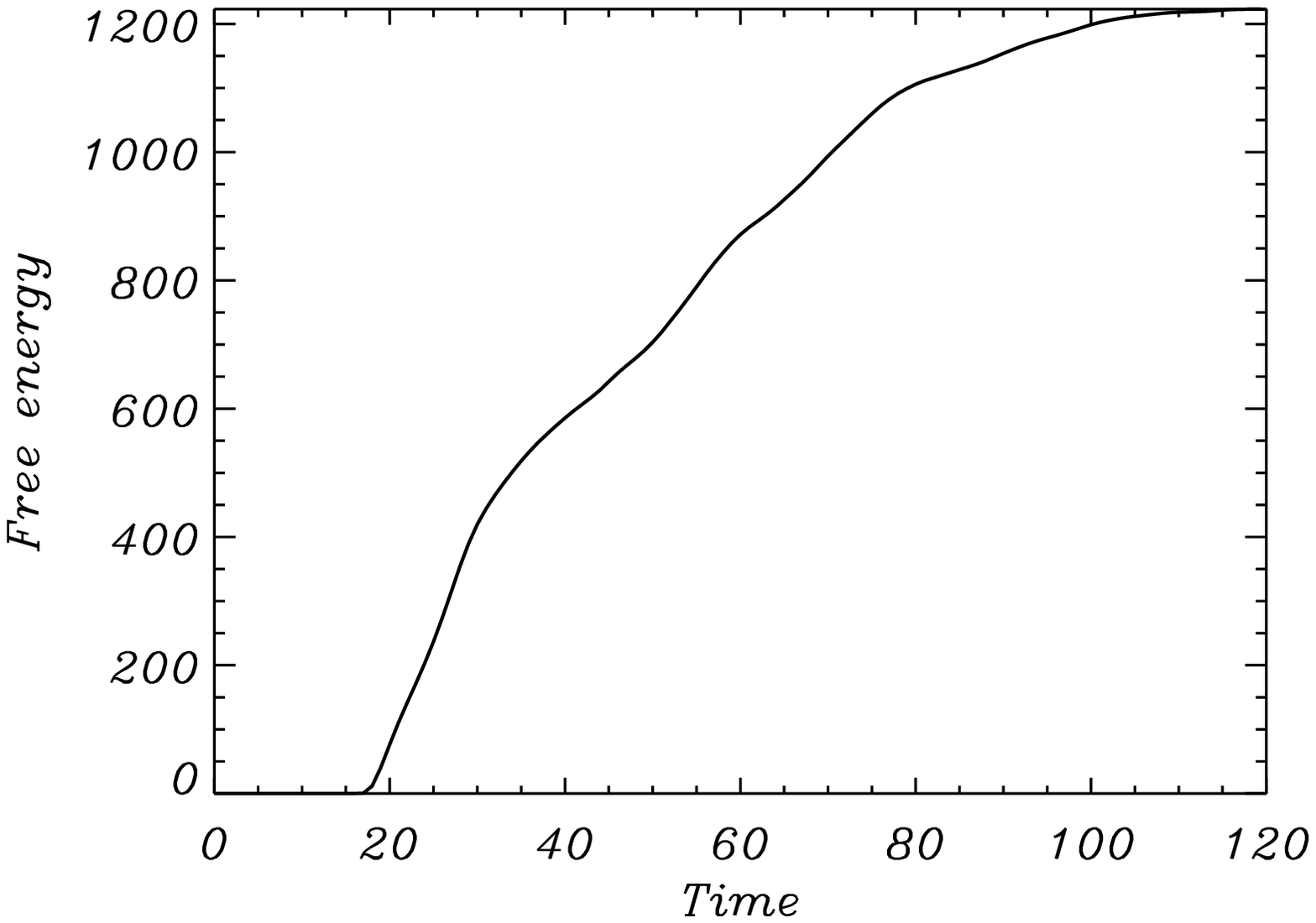}
\subcaption{$E_{\text{free}}$}
\label{subfig:freeenergyz0}
\end{subfigure}\\ 
\begin{subfigure}{0.48\textwidth}
\centering
\includegraphics[scale=0.38]{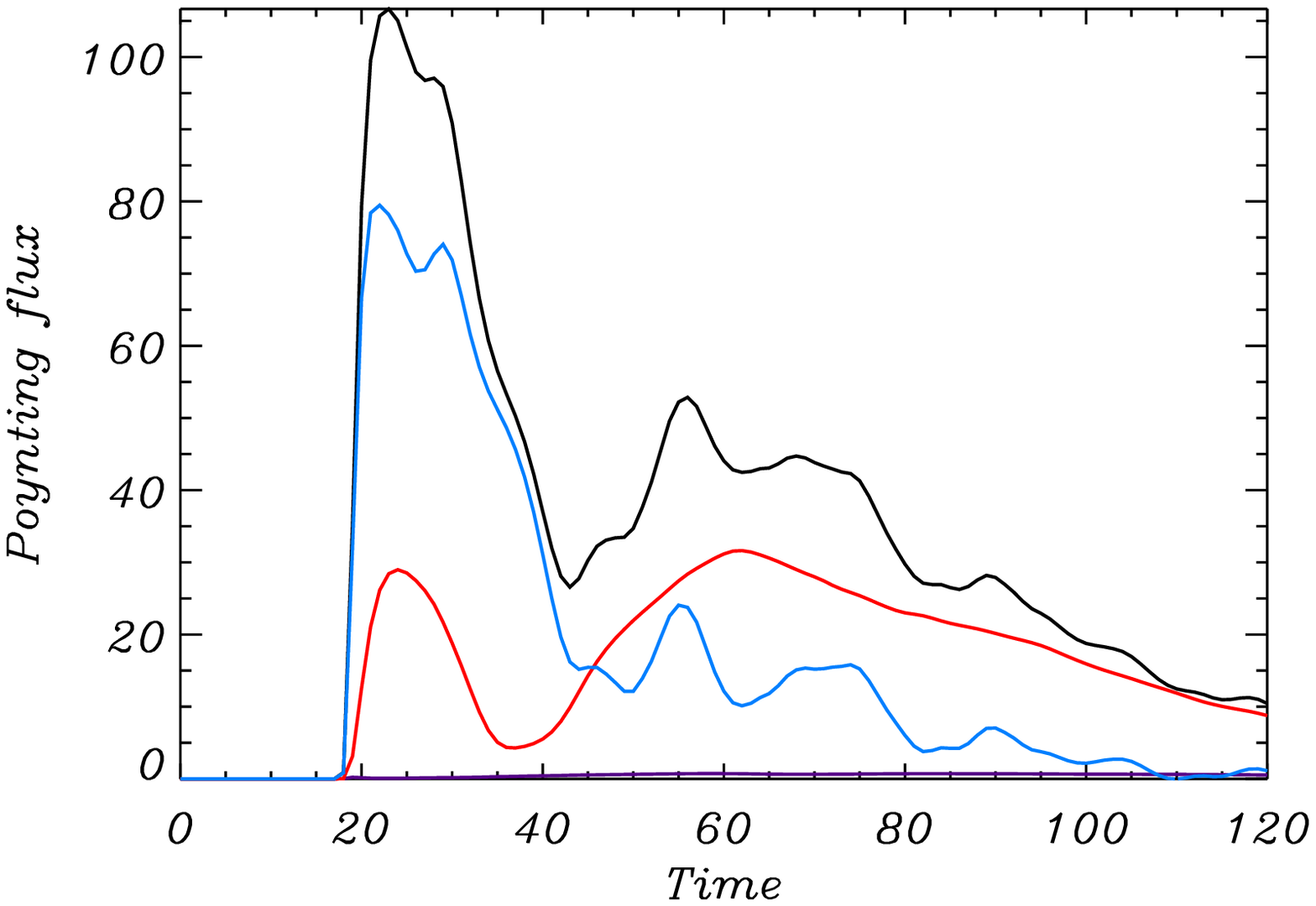}
\subcaption{$F_{\text{P}}$}
\end{subfigure}
\caption{Evolution of (a) the free energy when calculated above $z=0$ and (b) the vertical Poynting flux through the surface $z=0$. The total Poynting flux is split into the \emph{emergence} term (blue), the \emph{shear} term (red) and the \emph{resistive} term (purple) as defined in equation~\eqref{eq:pflux}.}
\label{fig:freeenergyz0}
\end{figure}

Now that we have demonstrated the transport of magnetic helicity from the interior portion of the domain to the coronal portion, we consider the magnetic energy and its distribution across the domain. In order to understand the amount of magnetic energy available for solar eruptive events, we consider the free magnetic energy relative to the potential field. That is, we calculate the excess magnetic energy after subtracting off the energy stored in the potential field, i.e. $\displaystyle{E_{\text{free}} = \int{\mathbf{B}^2/2~\mathrm{d}V} - \int{\mathbf{B}_p^2/2~\mathrm{d}V}}$. The evolution of the free magnetic energy above $z=0$ is shown in Figure~\ref{subfig:freeenergyz0}.
The excess energy above $z=0$ builds from the time the field first emerges. To investigate the contributions to magnetic energy by flux through the photospheric boundary we consider the Poynting flux through $z=0$ as given by,
\begin{equation}
F_{P} = \int_{z=0}{B^2 v_z~\mathrm{d}x\mathrm{d}y}-\int_{z=0}{(\mathbf{v \cdot B}) B_z~\mathrm{d}x\mathrm{d}y} + \eta\int_{z=0}{(\mathbf{j} \times \mathbf{B})_k~\mathrm{d}x\mathrm{d}y}.\label{eq:pflux}
\end{equation}
The first term contributing to the Poynting flux in equation~\eqref{eq:pflux} corresponds to the contribution by vertical flows owing to emergence, the second denotes the generation of magnetic energy by shearing/rotational flows and the third term is a result of resistive effects. The rate of increase of energy is largest during the initial stages of emergence due primarily to the emergence term. However, later the shear term (attributed to by rotational motions) is the main contributor to magnetic energy increase in the atmosphere. This pattern corroborates the trend that appeared in the helicity flux whereby vertical flows dominate the flux initially and horizontal flows become important later. Keeping in mind the precaution in Section~\ref{sec:hel} about the individual terms of helicity flux, the behaviour of the Poynting flux helps us to believe that the helicity flux trend may have physical meaning. At the end of the experiment, the free magnetic energy transported to the atmosphere has reached $8.2 \times 10^{22}$~J ($8.2 \times 10^{29}$ ergs).

\subsection{Force-free parameter}

\begin{figure}[!h]
\centering
\begin{subfigure}{0.48\textwidth}
\centering
\includegraphics[scale=0.28]{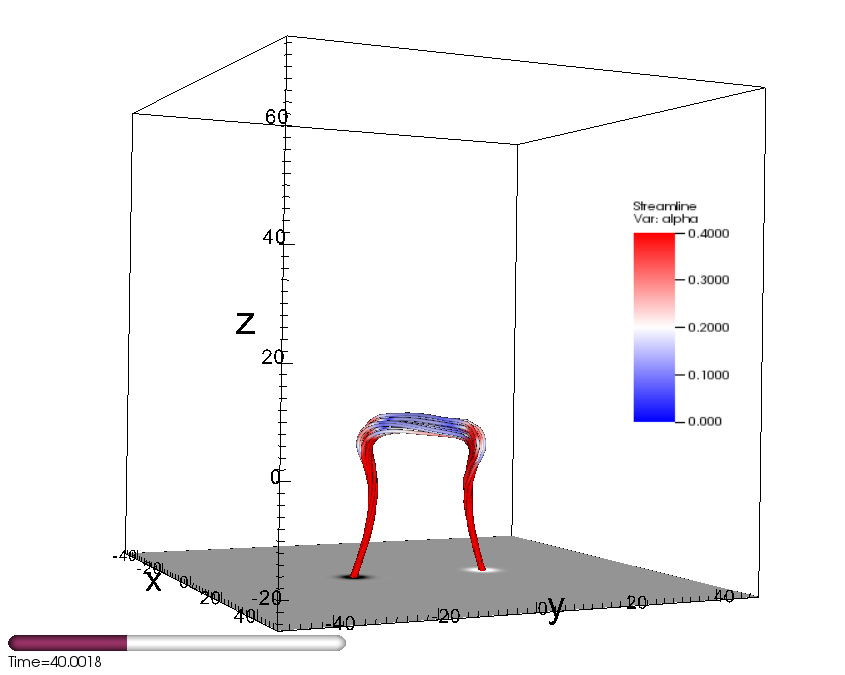}
\subcaption{$t=40$.}
\end{subfigure}\\
\begin{subfigure}{0.48\textwidth}
\centering
\includegraphics[scale=0.28]{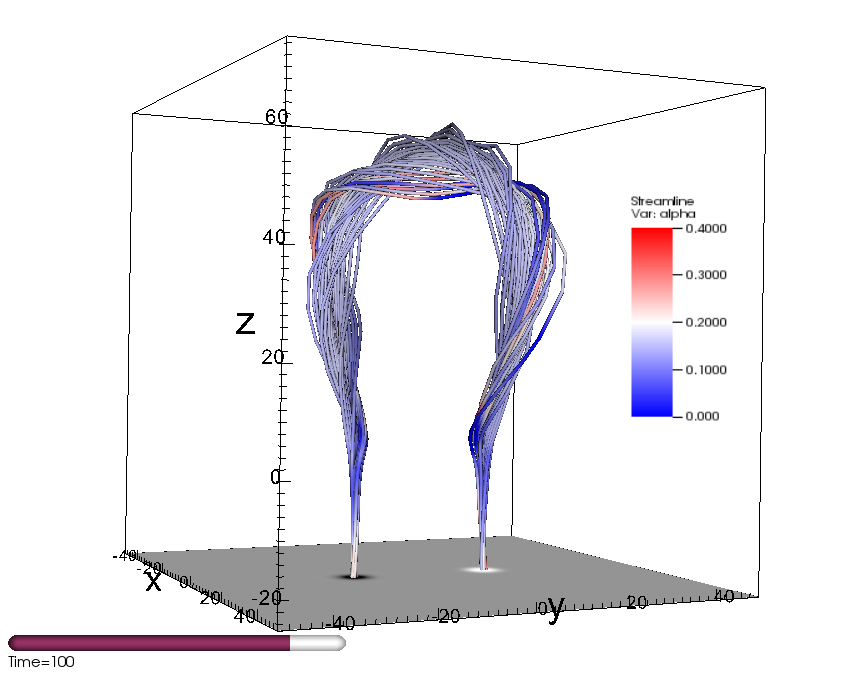}
\subcaption{$t=100$.}
\end{subfigure}\\
\begin{subfigure}{0.48\textwidth}
\centering
\includegraphics[scale=0.24]{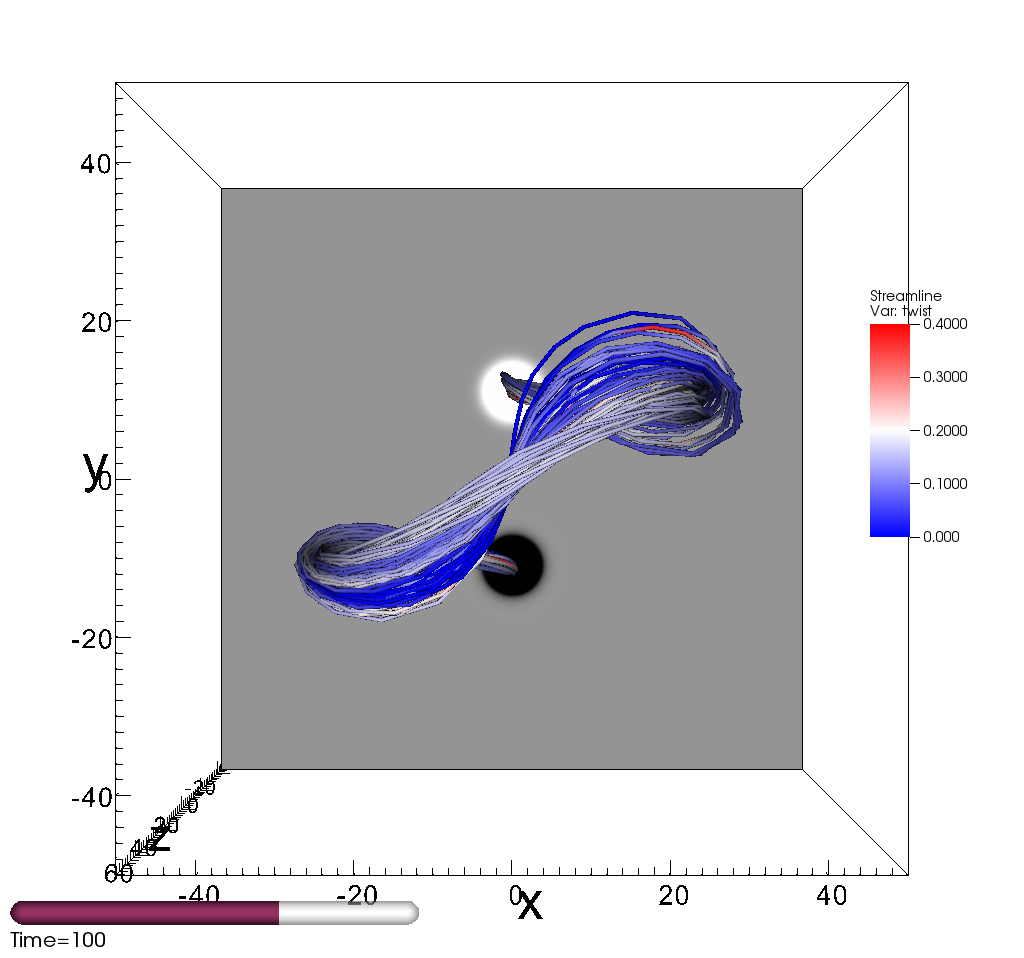}
\subcaption{$t=100$.}
\label{subfig:twistedS}
\end{subfigure}
\caption{Visualisation of magnetic fieldlines traced from both footpoints coloured by the parameter $\alpha$ such that red represents a strong twist ($0.2 < \alpha < 0.4$) and blue denotes a weaker twist ($0 < \alpha < 0.2$).}
\label{fig:rottwistmovie}
\end{figure}

Now that we have considered the behaviour of the plasma flows, current and magnetic helicity and energy, a proxy for the local twist is presented. Consider the quantity, $\alpha$, normally referred to as the force-free parameter or sometimes the fieldline torsion parameter,
\begin{equation}
\alpha = (\mathbf{\nabla} \times \mathbf{B}) \cdot \mathbf{B}/B^2.\nonumber 
\end{equation}

In order to investigate this expression, we trace $\alpha$ along fieldlines as a tool to help us visualise the distribution of twist along the field as shown in Figure~\ref{fig:rottwistmovie}. Initially the field is coloured red to signify the field is highly twisted with a value of $\alpha$ greater than $0.2$. Rapid emergence results in a coronal magnetic flux that is initially quite untwisted~\citep{Longcope2000} as shown by the low coronal $\alpha$. A clear gradient in $\alpha$ develops along the fieldlines from a high magnitude $\alpha$ in the interior to a lower magnitude of $\alpha$ in the atmosphere as displayed in Figure~\ref{fig:rottwistmovie}.~\cite{Fan2009} suggests that this gradient produces a torque that drives the rotational motions observed. This result has been proven by~\cite{Longcope1997} and it is suggested that this motion will continue until the gradient in $\alpha$ is removed~\citep{Longcope2000}. This agrees with our results.

Later, after the rotational motions have set in at the photosphere, the interior field is left coloured blue with weak twist ($\alpha < 0.2$) and the atmospheric field is threaded with a mixture of blue, red and white fieldlines. Thus, we can conclude that the field in the legs of the tube undergo a definitive untwisting motion as the field is transformed from highly twisted to weakly twisted by the end of the simulation. Although a large portion of the atmospheric field is coloured blue due to the expansion and stretching of the field, some fieldlines are coloured red indicating that some highly twisted field has been transported into the atmosphere. This is demonstrated in Figure~\ref{subfig:twistedS} where the twisted structure of the atmospheric field is shown. As $\displaystyle{\alpha \sim 1/L}$, an increase in the length scale results in a decrease in $\alpha$. The length scales in the corona are much larger so a decrease in $\alpha$ in the atmosphere can be explained by the expansion and stretching of magnetic fieldlines. The length scales in the interior, on the other hand, are much smaller and so we explain a decrease in $\alpha$ by an untwisting of the field.

\subsection{Propagation of torsional Alfv\'{e}n wave}
As touched on earlier, our results indicate that the rotational motions we observe may be governed by some form of torsional Alfv\'{e}n wave propagating downwards as a consequence of the transport of twist from the interior to the corona. The travel time for an Alfv\'en wave to propagate from the photosphere to the base of the domain is approximately $20$ time steps. This suggests that an Alfv\'en wave would take approximately $40$ time steps to travel down to the base, reflect and return to the photospheric plane. We propose that this downward propagating Alfv\'en wave, first proposed by~\cite{Longcope2000}, untwists the magnetic field in the interior. The rotation will only slow down once the reflected wave has returned to the photosphere. This appears to be in fairly good agreement with Figure~\ref{fig:rotavevortz0} where the rapid rotation and large $|\omega_z|$ occurs from about $t=50$ to $t=90$.

\section{Conclusions/Discussions}
\label{sec:conclude}

We have presented a $3$D MHD numerical experiment of the emergence of a toroidal flux tube from the solar interior through the photosphere and into the solar corona. Based on our detailed analysis, there is evidence that the interior magnetic field untwists and the photospheric footprints undergo a rotation. This rotational motion acts to untwist the interior field fixed at the base and injects twist into the emerged atmospheric portion. Our analysis of the plasma vorticity at the photospheric plane reveals that significant vortical motions develop in the centre of the sunspots. A definitive rotation of the sunspots is also demonstrated by tracking the fieldlines and calculating the rotation rate of the fieldlines threading the sunspot. Rotations of the order of one full rotation ($360\degree$) are observed in our experiment. This is similar in magnitude to the angles of rotation reported in studies that concluded a direct relationship between swift sunspot rotation and enhanced eruptive activity (\cite{Brown2003},~\cite{Yan2007} and~\cite{Yan2009} etc.). However, the sunspots in our experiment rotate by one full rotation over the course of about $40$ minutes, whereas~\cite{Brown2003} found sunspots to rotate through angles of the order of $200\degree$ over a period of days. The timescales in this experiment are clearly not in line with observations. However, this is related to the size of our emerging active region ($\sim10$Mm). If we scale up our experiment to a more typical active region, we expect the timescales to be much larger and in line with observations. This requires further investigation.

The direct emergence of flux paired with the continual rotational motions at the photosphere transport  magnetic energy and helicity from the solar interior to the corona. The magnetic energy in the atmosphere reaches a value of $8.2 \times 10^{22}~$J by the latter stage of the experiment. The Poynting flux of energy is split into contributions from horizontal shearing and vertical emergence terms. The initial flux of energy across the photospheric boundary is dominated by emergence but latterly the primary source is the horizontal shear. The rate of change of relative magnetic helicity in the solar atmosphere also has two predominant sources; namely helicity flux due to emergence and helicity flux by rotational motions. In a similar trend to the energy, initially the predominant source of helicity is the emergence of the magnetic flux tube but later this is replaced by the flux of helicity due to rotational motions at the photospheric level. The magnetic helicity transported to the atmosphere reaches a value of $3.6 \times 10^{19}$Mx$^2$. As well as the production of helicity in the atmosphere, we find a clear decrease in the magnetic helicity in the interior, supporting our understanding that this portion of the field undergoes an untwisting motion as also evidenced by a clear decrease in $j_z$ in the interior. 

The appearance of rotational motions centred on both sunspots has been found before in other MHD simulations including~\cite{Magara2006} and~\cite{Fan2009}.~\cite{Fan2009} also explained these rotations as a consequence of torsional Alfv\'{e}n wave propagation and established an increase in helicity in the atmosphere. Our work, however, explicitly discusses the effect that this rotational motion has on the interior portion of the field by establishing a depletion in the magnetic helicity stored in the interior segment of the domain and a drop in the vertical current in this region. We also show that the magnetic tension force may govern this rotational motion as it appears to produce an unbalanced torque that drives the rotation. Furthermore, we explicitly rule out that the rotation observed may be an apparent effect, helping us to explain the rotation primarily as a dynamical result of the emergence of magnetic flux. In addition, we trace fieldlines from the base of the domain as they pass through the photosphere and explicitly calculate their angles of rotation which appear to be approximately in line with the angles calculated in observations.  By considering the trajectories of these selected fieldlines, we find a remarkable visual representation of two fieldlines rotating in a circular motion around the axis (the centre of the sunspot), as shown in Figure~\ref{fig:rotfieldtrajectoryb}.

The simple hypothesis from~\cite{Longcope2000} has been confirmed in this 3D resistive MHD model, namely that only a fraction of the current carried by a twisted flux tube will pass into the corona and that the propagation of a torsional Alfv\'{e}n wave at the time of emergence will transport twist from the highly twisted interior to the stretched coronal loops. The rotational motions we observe are a manifestation of the propagation of these waves.

In a future paper we plan to perform a parameter study where we investigate the effects of the field strength $B_0$ and twist $\alpha$ on our analysis of the rotation of sunspots. This will allow us to determine if there is a relationship between the strength or twist of the sub-photospheric flux tube and the amount of vorticity in the sunspots or the rate at which the sunspots rotate. Another possible avenue for future research is to add in an ambient magnetic field to the coronal portion of the simulation domain. This would add in a further realism and allow us to understand the effect of an ambient field to the rate of rotation, as well as understanding whether rotation can lead to an eruption in our experiments.

Furthermore, we would like to try and understand why the rotation rate we calculate in our simulation is much larger than those calculated in observations. There are many possibilities for this discrepancy, including the size of active region which was discussed earlier. In addition, varying the strength or twist of the tube may change the time it takes for the flux tube to rise to the photosphere and hence govern the rotation rate of the tube. The model presented by~\cite{Longcope2000} predicts that the level of rotation will depend on the rapidity of flux emergence so we plan to investigate how this affects the rotation. The length of time for the rotation may also be related to the depth at which the flux tube is anchored; this is another approach that requires investigation.

\begin{acknowledgements}
 ZS acknowledges the financial support of the Carnegie Trust for Scotland and CMM the support of the Royal Society of Edinburgh. This work used the DIRAC 1, UKMHD Consortium machine at the University of St Andrews  and the DiRAC Data Centric system at Durham University, operated by the Institute for Computational Cosmology on behalf of the STFC DiRAC HPC Facility (www.dirac.ac.uk). This equipment was funded by BIS National E-infrastructure capital grant ST/K00042X/1, STFC capital grant ST/H008519/1, and STFC DiRAC Operations grant ST/K003267/1 and Durham University. DiRAC is part of the National E-Infrastructure. The authors are grateful to Dr. M. Berger for useful comments concerning magnetic helicity and very thankful to Dr. A. Wright for his helpful discussions and input concerning the propagation of a torsional Alfv\'en wave. We also thank the unknown referee for the helpful comments.
\end{acknowledgements}
\bibliographystyle{aa}
\bibliography{MyCollection}

\end{document}